**TITLE:**
*Go Figure: Transparency in neuroscience images preserves context and clarifies interpretation*

**Authors:**
1. Paul A. Taylor, Scientific and Statistical Computing Core, NIMH, NIH, Bethesda, MD, USA
2. Himanshu Aggarwal, Inria, CEA, Université Paris-Saclay, Palaiseau, 91120, France
3. Peter A. Bandettini, Section on Functional Imaging Methods, NIMH, NIH, Bethesda, MD, USA
4. Marco Barilari, Crossmodal Perception and Plasticity Lab, Institute of Neuroscience (IoNS) and Institute of Research in Psychology (IPSY), Université Catholique de Louvain, 1348 Louvain-la-Neuve, Belgium
5. Molly G. Bright, Department of Physical Therapy and Human Movement Sciences, Feinberg School of Medicine, Northwestern University, Chicago, IL, USA; Department of Biomedical Engineering, McCormick School of Engineering and Applied Sciences, Northwestern University, Evanston, IL, USA
6. César Caballero-Gaudes, Basque Center on Cognition, Brain and Language, San Sebastian-Donostia, Spain; Ikerbasque, Basque Foundation for Science, Bilbao, Spain
7. Vince D. Calhoun, Tri-institutional Center for Translational Research in Neuroimaging and Data Science (TReNDS), Georgia State, Georgia Tech, Emory, Atlanta, GA, USA
8. Mallar Chakravarty, Cerebral Imaging Centre, Douglas Mental Health University Institute, Montreal, QC, Canada; Department of Psychiatry, McGill University, Montreal, QC, Canada; Department of Biomedical Engineering, McGill University, Montreal, QC, Canada
9. Gabriel A. Devenyi, Cerebral Imaging Centre, Douglas Mental Health University Institute, Montreal, QC, Canada; Department of Psychiatry, McGill University, Montreal, QC, Canada
10. Jennifer W. Evans, Experimental Therapeutics and Pathophysiology Branch, NIMH, NIH, Bethesda, MD, USA
11. Eduardo A. Garza-Villarreal, Department of Behavioral and Cognitive Neurobiology, Institute of Neurobiology, Universidad Nacional Autónoma de México campus Juriquilla, Querétaro, Mexico
12. Jalil Rasgado-Toledo, Department of Behavioral and Cognitive Neurobiology, Institute of Neurobiology, Universidad Nacional Autónoma de México campus Juriquilla, Querétaro, Mexico
13. Rémi Gau, Inria, CEA, Université Paris-Saclay, Palaiseau, 91120, France
14. Daniel R. Glen, Scientific and Statistical Computing Core, NIMH, NIH, Bethesda, MD, USA
15. Rainer Goebel, Department of Cognitive Neuroscience, FPN, Maastricht University, Maastricht, NL;  Brain Innovation, Maastricht, NL
16. Javier Gonzalez-Castillo, Section on Functional Imaging Methods, NIMH, NIH, Bethesda, MD, USA; Basque Center on Cognition, Brain and Language, San Sebastian-Donostia, Spain
17. Omer Faruk Gulban, Department of Cognitive Neuroscience, FPN, Maastricht University, Maastricht, NL;  Brain Innovation, Maastricht, NL
18. Yaroslav Halchenko, Department of Psychological and Brain Sciences, Dartmouth College, Hanover, NH, USA

19. Daniel A. Handwerker, Section on Functional Imaging Methods, NIMH, NIH, Bethesda, MD, USA
20. Taylor Hanayik, Wellcome Centre for Integrative Neuroimaging, FMRIB, University of Oxford, Oxford, UK
21. Peter D. Lauren, Scientific and Statistical Computing Core, NIMH, NIH, Bethesda, MD, USA
22. David A. Leopold, Systems Neurodevelopment Laboratory, NIMH, NIH, Bethesda, MD, USA
23. Jason P. Lerch, Wellcome Centre for Integrative Neuroimaging, Nuffield Department of Clinical Neurosciences, University of Oxford, Oxford, UK; Department of Medical Biophysics, University of Toronto, Toronto, ON, Canada
24. Christian Mathys, Institute of Radiology and Neuroradiology, Evangelisches Krankenhaus Oldenburg, Universitätsmedizin Oldenburg, Oldenburg, Germany; Research Center Neurosensory Science, Carl von Ossietzky Universität Oldenburg, Oldenburg, Germany
25. Paul McCarthy, Wellcome Centre for Integrative Neuroimaging, FMRIB, Nuffield Department of Clinical Neurosciences, University of Oxford, UK
26. Anke McLeod, Department of Radiology and Nuclear Medicine, University Hospital Magdeburg, Otto-Von-Guericke University, Magdeburg, Germany
27. Amanda Mejia, Department of Statistics, Indiana University, Bloomington, USA
28. Stefano Moia, Department of Cognitive Neuroscience, FPN, Maastricht University, Maastricht, NL
29. Thomas E. Nichols, Big Data Institute, Li Ka Shing Centre for Health Information and Discovery, Nuffield Department of Population Health, University of Oxford, UK; Wellcome Centre for Integrative Neuroimaging, FMRIB, Nuffield Department of Clinical Neurosciences, University of Oxford, UK
30. Cyril Pernet, Neurobiology Research Unit, Rigshospitalet, Denmark
31. Luiz Pessoa, Department of Psychology, University of Maryland, College Park, MD, USA
32. Bettina Pfleiderer, Clinic of Radiology, Medical Faculty, University of Münster, Münster, Germany
33. Justin K. Rajendra, Scientific and Statistical Computing Core, NIMH, NIH, Bethesda, MD, USA
34. Laura D. Reyes, Laboratory on Quantitative Medical Imaging, National Institute of Biomedical Imaging and Bioengineering, NIH, MD, USA
35. Richard C. Reynolds, Scientific and Statistical Computing Core, NIMH, NIH, Bethesda, MD, USA
36. Vinai Roopchansingh, Functional MRI Facility, NIMH, NIH, Bethesda, MD, USA
37. Chris Rorden, McCausland Center for Brain Imaging, Department of Psychology, University of South Carolina, Columbia, SC 29208, USA
38. Brian E. Russ, Center for Biomedical Imaging and Neuromodulation, Nathan Kline Institute, 140 Old Orangeburg Road, Orangeburg, NY 10962, USA;  Nash Family Department of Neuroscience and Friedman Brain Institute, Icahn School of Medicine at Mount Sinai, One Gustave L. Levy Place, New York, NY 10029, USA; Department of Psychiatry, New York University at Langone, One, 8, Park Ave, New York, NY 10016, USA
39. Benedikt Sundermann, Institute of Radiology and Neuroradiology, Evangelisches Krankenhaus Oldenburg, Universitätsmedizin Oldenburg, Oldenburg, Germany; Research

Center Neurosensory Science, Carl von Ossietzky Universität Oldenburg, Oldenburg, Germany; Clinic of Radiology, Medical Faculty, University of Münster, Münster, Germany
40. Bertrand Thirion, Inria, CEA, Université Paris Saclay, France.
41. Salvatore Torrisi, University of California, San Francisco, Department of Radiology & Biomedical Imaging, SF CA; San Francisco Veteran Affairs Health Care System, SF CA
42. Gang Chen, Scientific and Statistical Computing Core, NIMH, NIH, Bethesda, MD, USA

**ABSTRACT**

Visualizations are vital for communicating scientific results. Historically, neuroimaging figures have only depicted regions that surpass a given statistical threshold. This practice substantially biases interpretation of the results and subsequent meta-analyses, particularly towards non-reproducibility. Here we advocate for a "transparent thresholding" approach that not only highlights statistically significant regions but also includes subthreshold locations, which provide key experimental context. This balances the dual needs of distilling modeling results and enabling informed interpretations for modern neuroimaging. We present four examples that demonstrate the many benefits of transparent thresholding, including: removing ambiguity, decreasing hypersensitivity to non-physiological features, catching potential artifacts, improving cross-study comparisons, reducing non-reproducibility biases, and clarifying interpretations. We also demonstrate the many software packages that implement transparent thresholding, several of which were added or streamlined recently as part of this work. A point-counterpoint discussion addresses issues with thresholding raised in real conversations with researchers in the field. We hope that by showing how transparent thresholding can drastically improve the interpretation (and reproducibility) of neuroimaging findings, more researchers will adopt this method.

**INTRODUCTION**

Beyond performing experiments and recording data, a scientist is responsible for synthesizing information, interpreting it within the context of prior knowledge, and communicating it effectively. This curation and distillation involves a careful balance of contextualization and concise communication, without losing meaningful information. Data visualization is itself an important analysis step with key processing choices to be made, though their impact is often overlooked and underappreciated. Here we examine these aspects for results reporting in neuroimaging, where typical datasets are complex and multi-dimensional. We start by describing how the field and its research questions have changed over time, and then discuss how the presentation of results in brain images should similarly evolve. The proposed improvements are straightforward and implementable in a large number of widely used software packages. Many of the presented examples use functional magnetic resonance imaging (FMRI) data, but the concepts and methods of improving results reporting are directly applicable to other imaging modalities.

*Background: historical context*

Localization was an early focus of neuroimaging researchers, who broadly conceptualized the brain as a set of discrete regions with well-defined functions to be identified and displayed, in line with empirical procedures from lesion based neuropsychology and early cognitive psychology (e.g., see Savoy, 2001). Unfortunately, neuroimaging signals—particularly FMRI recordings—are noisy, dynamic, and contain smooth patterns whose size and shape is hard to delineate. To combat these initial challenges, neuroscientists turned their attention to standard null hypothesis significance testing as a way to implement a clear filtering mechanism. Within this framework, strict thresholding was a key processing step for localizing a small number of regions of interest and reducing potential false positives (Forman et al., 1995; Nichols and Hayasaka, 2003; Smith and Nichols, 2009). An emphasis was placed on the results having a small number of suprathreshold clusters and rigorous boundaries of statistical significance. Subthreshold regions were interpreted as simply noise or non-neuronal features (negative-signed activations were even filtered in many cases, prior to interest in task-negative networks). Therefore, researchers presented results with strict or "opaque" thresholding: wherever a statistic value was subthreshold, all modeling results—including the effect estimate, *p*-value and statistic itself—are fully hidden from view and withheld from consideration.

However, this approach comes with inherent challenges. The adoption of univariate statistics on interdependent data like voxels and vertices requires statistical adjustment for multiple comparisons to reduce false positives across these thousands of tests. At the same time, this also causes meaningful effects to fall below the substantially decreased statistical power (Cremers et al., 2017; Lohmann et al., 2017; Bacchetti, 2013), causing a "tip of the iceberg effect" (Pang et al., 2023; Noble et al., 2024; Sundermann et al., 2024). Adopting multivariate analyses (e.g. searchlight-based multivariate pattern analysis, MVPA) does not solve this problem, since statistical peaks might not correspond to biologically meaningful locations, but instead to the center of maximally informative neighborhoods (Etzel et al., 2013). Consequently, replication studies using similarly stringent thresholds may yield results that appear strikingly different, potentially causing undue concerns about low reliability and highlighting the limitations of this traditional framework. Even within a single study, relevant results may fall below the strict corrective adjustments (i.e., fall "below the waterline"), biasing the evaluation and interpretation.

Over time, methodological and conceptual changes have also occurred in the field. Modern neuroimaging presents a more intricate and nuanced picture of brain activity. Network-based and connectomic studies have become much more common, shifting to a paradigm in which most functions involve the interaction of many parts of the brain to varying degrees (e.g., Calhoun et al, 2001; Calhoun et al., 2002; Greicius et al., 2003; Damoiseaux et al., 2006; Hagmann et al., 2008; Fair et al., 2009; Smith et al., 2009; Biswal et al., 2010; Van Essen et al., 2013; Pessoa, 2014). While regions with high statistical strength might still be particularly important, they do not simply indicate small regions turning on/off in isolation. Responses modulate, and importance also lies with other parts of their (or other) networks that might have weaker effects (e.g., see Noble et al. 2024). That is, even when focused on "significant clusters," the subthreshold results across the rest of the brain will still provide necessary context for understanding their role and for having a more complete picture of how the brain is behaving. Gonzalez-Castillo et al. (2012)'s deep scanning study further showed brain activation at many scales and changing extents of regional activation with increased data. Recent approaches of modeling eigenmodes of brainwide function have reinforced the importance of visualizing nonlocal and subthreshold effects (Pang et al., 2023).

As a result, the idea of an exactly zero response in any gray matter seems unlikely. Treating it as such—which is what standard opaque thresholding effectively does—creates statistical issues (Cremers et al., 2017). Even if zero effects existed, concluding that subthreshold test statistics imply zero effect amounts to confirmation of the null hypothesis, which goes against the principles of hypothesis testing. In the view of modern neuroimaging, opaque thresholding also wastes meaningful information, since responses and effects are not simply localizable in an on/off manner, as responses occupy a continuous spectrum (Chen et al., 2022). This is one reason that clinical practitioners often include fully unthresholded images in their assessments, to see more context and reduce false negatives (Voets et al., 2025). In general, the results of a neuroimaging study should account for the non-dichotomous nature of data by including subthreshold information, both when the study authors are interpreting it and when readers are engaging with their work.

*Reporting results: the past and the future*

The issue of how figures are made might seem like merely a stylistic choice, but it is central to how scientists evaluate and interpret results, how readers assess them, and how meta-analyses compare them. Data visualization is an analysis step, and thresholding data is one of the final processing choices researchers make in a study. It has important consequences for evaluating and understanding results.

The current standard practice of opaque thresholding is rooted in the assumptions of the earliest neuroimaging studies. It has remained largely unchanged and, as a consequence, so have the basic figures and representations of results in neuroimaging studies, even though our understanding of brain function has grown in many ways. Previous work has noted problems with opaque thresholding (Allen et al., 2012; Chen et al., 2022; Taylor et al., 2023; Sundermann et al., 2024). Motivated by this, here we identify and focus on three ways that opaque thresholding negatively impacts the fundamental interpretations and comparisons of neuroimaging results:
    1) **Unrealistic biology:** Opaque thresholding treats all subthreshold regions as if they had zero effect rather than simply statistically weaker observations. This creates an unrealistic ON/OFF picture of localized effects, and is not consistent with the current understanding of brain functioning.
    2) **Ambiguity:** Brain regions are generally part of overlapping networks, rather than purely isolated and independent. Opaque thresholding removes the context of any results: how quickly effects drop off spatially, what network(s) a cluster is involved with, etc. Other parts of the brain might have higher uncertainty but they still contain useful context, such as evidence for the full range of effects, wider network interpretations, and comparisons.
    3) **Bias:** The thresholding of a continuous brain effect or related metric mathematically introduces biases and hypersensitivity to small (often arbitrary) differences, such as between

effects minimally below and above the threshold. These will negatively impact within-study evaluation and cross-study reproducibility.

As a consequence, opaque thresholding undermines the content of the results themselves. It compromises the ability to make holistic and accurate interpretations by study authors, readers, and meta-analyses at the most fundamental levels.

To improve neuroimaging visualization, Allen et al. (2012) proposed transparent thresholding as an effective way to include *both* suprathreshold *and* subthreshold information in reported results. This is a simple, meaningful, and straightforward solution whereby suprathreshold regions are highlighted in an image by being opaque and outlined, while subthreshold results are also included by using an opacity that decreases with their absolute statistical value. This balanced approach allows for having a concise summary of regions with the strongest effects (the same suprathreshold regions in existing figures), together with a graded assessment of activity across the rest of the brain (reporting information across the "missing majority" of the data that had been gathered and analyzed). By shifting to transparent thresholding, the data modeling and results are more accurately represented, *no information is lost*, and meaningful evidence is gained.

To date, this visualization approach has not been widely adopted in the field, but it has been applied effectively in several neuroimaging studies (a sample of these are provided in Table S1 of the Supplements). Its utility in providing more complete evaluations of results has also been shown in direct comparisons with standard, opaque thresholding. For example, Taylor et al. (2023) used transparent thresholding to reveal a previously underappreciated degree of consistency and reproducibility in the FMRI results of the Neuroimaging Analysis Replication and Prediction Study (NARPS; Botvinik-Nezer et al., 2020).

Here, we provide four examples that demonstrate the many benefits of retaining context in results. Each highlights an important aspect of reporting and interpreting results that is notably improved with transparent thresholding, namely: improving the understanding of a study, reducing hypersensitivity to arbitrary features (sample size, many processing choices and more), avoiding problematically selective reporting, and enhancing meta-analyses. In the later examples, we also argue that, relatedly, effect estimates from modeling should also be visualized whenever available. These provide an additional source of important information in both figures and comparisons. In the Discussion, we address several considerations and concerns that researchers in the field have raised about transparent thresholding. Changes to conventions often face hesitation, and these points are important to consider. But the scientific costs of using opaque thresholding are demonstrably high, and the benefits of showing context with transparent thresholding are both clear and abundant.

We note that one barrier to adopting a new approach is its availability to researchers. We highlight that transparent thresholding is now widely available across a large number of neuroimaging software packages. While a small number of toolboxes have previously included transparent thresholding, several others have been recently implemented or significantly streamlined this approach during this project. These tools are listed (with example images) in the Discussion, and further details are contained in the Supplements.

**RESULTS**

We present four examples using real data that illustrate the importance of showing context in brain images. These beneficial features include: removing ambiguity in interpreting the data (Ex. 1); accurately comparing data and assessing differences, both visually and in more formal meta-analyses (Ex. 2); avoiding hypersensitivity to non-physiological features, such as sample size (Ex. 3); and reinforcing robust and accurate interpretations (Ex. 4).

*Example 1: context to reduce ambiguity*

Fig. 1A shows group results from a set of Flanker task-based FMRI data (see Chen, Pine, et al. 2022) presented in the conventional style. In this figure, data processing includes opaque thresholding at a family-wise error (FWE) rate of 5% (via voxelwise $p$ = 0.001 and cluster-correction N = 40 voxels). The displayed results only contain the lobes of a single cluster appearing in the right intraparietal sulcus. Since opaque thresholding is applied, no other information is conveyed for this slice than the fact that the absolute value of the statistic at every non-cluster location was subthreshold. Some potentially interesting clusters might be just one voxel below the estimated cut-off, but they remain entirely hidden. The reader can only interpret the biological implications of this study from this sparse statistical information alone: thus, only a single region appears to show significant response, and the activity seems fully lateralized in the right intraparietal sulcus.

The scale of information loss can be appreciated by viewing Fig. 1B, where the thresholding has been applied transparently to retain context in the same data. This reveals a brainwide set of nontrivial responses, many of which seem biologically relevant despite being statistically subthreshold. Note that the opacity fades quadratically with decreasing statistic value, so the most notable "extra" results are still near the threshold level (compare the colorbars in Figs. 1A-B). Seeing this context assists in interpreting the most significant regions, which are still highlighted with full opacity and outlines. It also reduces possible misinterpretation. For example, the richer context of the modeling results implies that the "lone lateralized cluster" actually has much stronger left-right symmetry than judged from Fig. 1A (though still with a larger response on the right). The arbitrary influence of opaque thresholding on laterality measures has been particularly noted in clinical contexts (Ruff et al., 2008; Seghier, 2008; Suarez et al., 2009), and though not yet widely adopted, transparent thresholding would greatly improve evaluations. We also see that opaque thresholding has over-reduced the results of the whole-brain modeling, while Fig. 1B contains a more accurate representation of the full study evidence and invites further research to explore the involvement of network components (several of which also appear focal and lateralized).

The interpretational consequences of over-reducing results are not just a quirk of FMRI, and cases from other fields provide useful lessons. For instance, in the classic example of Anscombe's Quartet (Anscombe, 1973), four scatterplots show data that have the exact same summary statistics and correlation value, but with very different underlying patterns (see Fig. 1C). If the results of the correlation analysis are only reported as summary statistics, important information is lost and one is highly likely to make an entirely incorrect inference about the data itself. Only by keeping the contextual information of the plot can the ambiguity of the summary values be resolved. Acknowledgment of these issues has led researchers in many areas to take action in improving their statistical reporting: they show scatterplots of data (not just fit values); they use violin plots and raincloud plots to detail distributions; and more.

Identical reasoning further emphasizes the importance of using transparent thresholding to retain context in neuroimaging reports. Fig. 1D shows images of four sets of possible results, related to the Flanker task in Fig. 1A. Transparent thresholding allows us to see that there are major differences among them, but with opaque thresholding they each reduce exactly to Fig. 1A and are indistinguishable. That is, opaque thresholding inherently produces problematic ambiguity because the interpretation of the statistically significant cluster would be dramatically different across the four cases, with each image representing a very different biological implication: 1) fairly symmetric left-right activation (actual response); 2) *anti*-symmetric left-right activation; 3) strongly right-lateralized response; 4) likely a noise- or artifact-driven outcome rather than a task-related physiological one. From the loss of context with opaque thresholding, there is little choice but to apply Occam's Razor and interpret the results as pointing toward strong or absolute laterality (something like Image #3 in Fig. 1D), which the fuller context does not support. By thresholding transparently, one observes a more complete representation of the results while still having the most significant regions highlighted.

Seeing results beyond the tip of the iceberg also improves localization, by revealing the spatial drop-off of the response (here, in terms of statistical value, but below we show how the effect evidence can be presented). This further provides useful information about the properties of the data itself such as its spatial smoothness, which reflects both acquisition and processing. Interestingly, spatial smoothness was highlighted as a major factor for varied outcomes in the NARPS project (Botvinik-Nezer et al., 2020), so having this information directly available in primary figures may be particularly useful for cross-study comparisons. Finally, subthreshold visualizations can help distinguish among potential underlying features, such as motion versus respiratory effects, additionally benefitting quality control and assurance efforts.

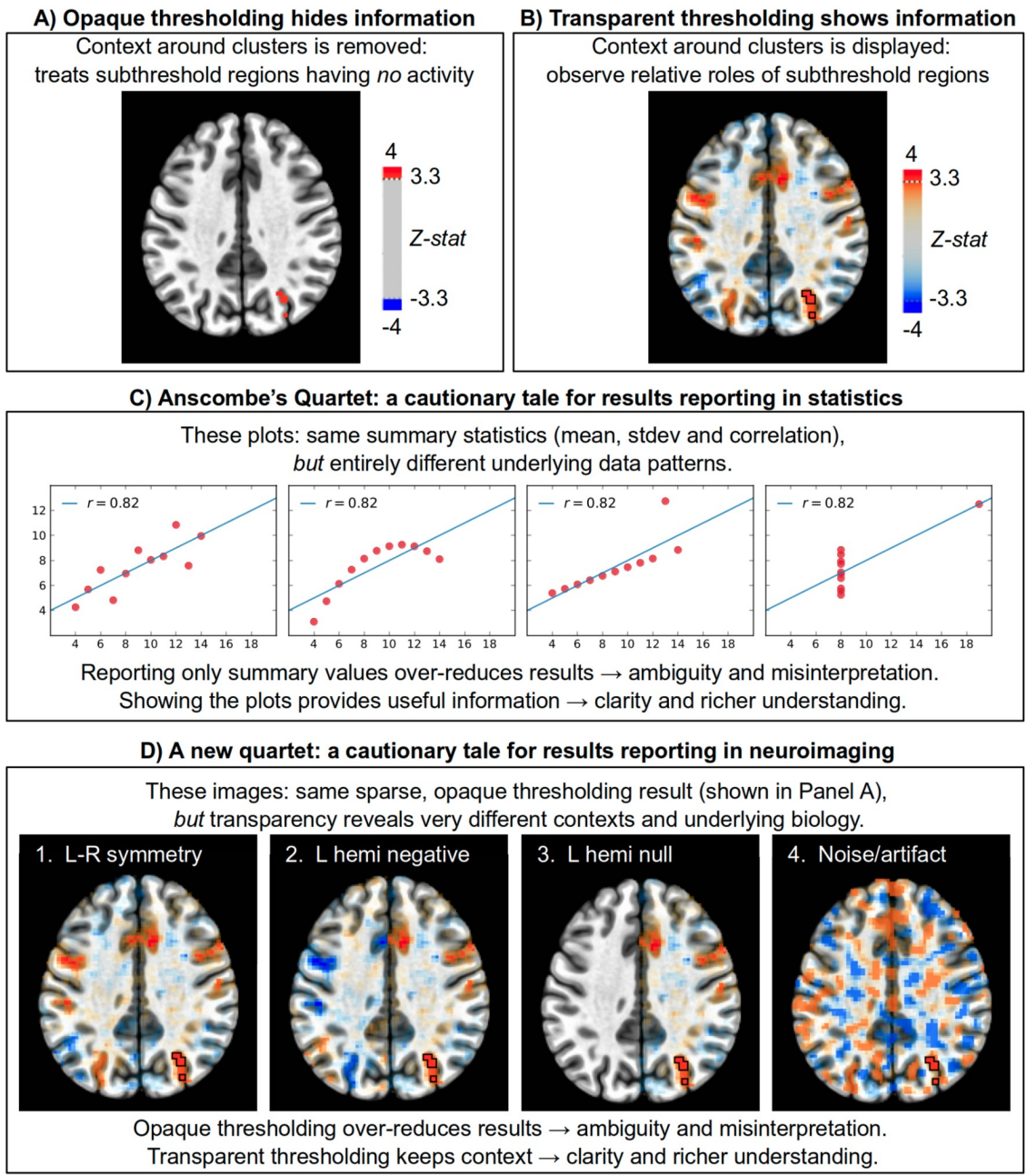

*Figure 1.* Results reporting examples, showing a single slice of task-based FMRI data (see Chen, Pine, et al., 2022). Each neuroimaging panel shows the same axial slice in MNI template space at z = 36S (image left = subject left), with thresholding is applied at voxelwise p = 0.001 and cluster size = 40 voxels (FWE = 5%). The data used for both overlay coloration and thresholding are the Z-score statistics. Panel A displays FMRI results using conventional strict (or opaque) thresholding, and shows one cluster in the right intraparietal sulcus. Panel B displays the same results with transparent thresholding (suprathreshold regions are opaque and outlined; subthreshold regions fade as the statistic decreases), revealing relevant context in the subthreshold regions that are hidden in A. Panel C shows a classic example from Anscombe (1973) of the risks of over-reducing data, here for a simple scatterplot. Panel D shows how the same considerations apply to neuroimaging: each dataset would have very different

*interpretations and biological implications, which can be appreciated with transparent thresholding (same colorbar as B), but when using opaque thresholding that context is lost and each slice reduces to the same image (that of panel A). Only by displaying the more full context with subthreshold visualization can the degeneracy be broken and results more accurately understood. Opaque thresholding removes context and can often lead to a misinterpretation of results.*

*Example 2: context to improve cross-study comparison and meta-analysis*

Retaining context is particularly important for comparing datasets and performing meta-analyses. The NARPS project gathered results from approximately 70 teams who had processed the same task-based FMRI data collection independently and reported on specific hypotheses. In their primary comparison of FMRI results, the NARPS authors assessed the similarity of binarized yes/no responses to nine region-specific hypotheses, based on opaquely thresholded statistical results. They reported observing a "substantial variability in reported binary results, with high levels of disagreement across teams on a majority of tested hypotheses." They also noted that the similarity of teams' opaquely thresholded maps, as measured by cluster overlap, was low. However, when they compared the *unthresholded* results, they found "a large cluster of teams had statistical maps that were strongly positively correlated with one another" and that "analyses of the underlying statistical parametric maps on which the hypothesis tests were based revealed greater consistency than expected from those inferences." That is, comparisons on results processed with standard thresholding showed relative disagreement, while those without thresholding showed a large amount of agreement. The second message has tended to be greatly underemphasized, and this widely cited paper has overwhelmingly been referenced simply as evidence of *high* variability across processing pipelines. Instead, it should be viewed as a demonstration of the influence thresholding has as a processing step and an important warning of the biases of opaque thresholding, which preferentially tilt study comparisons towards non-reproducibility.

We can see how the choice of thresholding explains the opposing meta-analytic results—high variability when thresholding vs surprising similarity without it—by visualizing a representative set of teams' results. Fig. 2A shows nine teams' results from analyzing the same FMRI dataset for NARPS Hypotheses 1, and applying opaque thresholding at |*Z*| or |*t*| = 3 (equivalent quantities due to the degree of freedom count; see Supplements). While a few of the teams share somewhat similar suprathreshold clusters, several contain almost no results (3rd column) or sparse regions, yielding an impression of high variability across the teams, consistent with the NARPS authors' primary meta-analysis findings. However, switching to transparent thresholding (Fig. 2B) immediately reveals that the statistical patterns are actually quite similar across the majority of teams, just with differing magnitudes. That is, the spatial patterns have high correlation, consistent with the secondary meta-analytic findings in NARPS. These observations are quantified and summarized in the similarity matrices shown in Fig. 2C and 2D, which are again consistent with NARPS's primary and secondary meta-analyses, respectively. Taylor et al. (2023) showed how these patterns of predominantly high similarity (with a small number of outliers) persist across the full set of teams' results and all hypotheses, when applying transparent thresholding.

This example demonstrates how having the full context in the images is key to understanding the full scope of results—particularly the *kind* of variability that is present.[1] The dominant variability across the NARPS teams' results is actually that of the *magnitude* of the statistics, rather than of the sign or spatial pattern of statistical maps. The opaquely thresholded maps cannot distinguish between these kinds of variability below an elevated cut-off point, and hence they bias the interpretation towards one of simply high variability and "lack of reproducibility." Seeing the full context as part of the comparison helps to reduce and resolve this bias.

Note that while opaque thresholding biases comparisons towards non-reproducibility, using transparent thresholding to retain context does *not* necessarily increase similarity. It merely allows for better informed comparisons. Consider the third column in each of Fig. 2A and 2B: while transparent thresholding reveals meaningful statistical patterns in the upper two images, it also shows that the bottom image values are uniformly quite low and negligible by comparison. This is another case of resolving ambiguity, as described in Fig. 1: what ostensibly appeared to be three examples of the same thing under opaque thresholding were actually revealed to be distinct by using transparency. Seeing more context in the bottom-middle image reveals that its results have much lower similarity to the others above it, which is also reflected in its much lower correlation value (Fig. 2D). In all cases, these more informative observations can lead to a more impartial assessment of a particular team's results, and encourage a helpful examination of the corresponding preprocessing and analytic choices.

Thus, while there is variability in the NARPS teams' processing and results, two different meta-analysis approaches provide very different assessments of it both visually and quantitatively. The one associated with opaque thresholding (which hides useful information, leaves ambiguity, and biases the comparison towards dissimilarity) suggests high variability. In contrast, the one associated with transparent thresholding (which leaves context, provides more modeling evidence, and represents a more complete picture of the data) suggests widespread similarity with varied magnitude. That is, *rather than being an inherent feature of the teams' results, the outcome of high variability is primarily due to the processing choice of applying opaque thresholding prior to comparison, which introduces a strong bias.* Instead, retaining context in the datasets improves both the mechanics and interpretation of the meta-analysis.[2]

---

1 It also shows the importance of including at least some images of the underlying study datasets themselves in comparisons report. No figures of individual teams' results were shown in the NARPS paper or its supplements, though the datasets were made publicly available.
2 We note that any formal meta-analysis here would be limited, since teams were only required to upload statistics data. This follows the unfortunately common practice that effect estimate information is rarely reported, even though it has many uses (Chen et al., 2017). With more detailed effect size data, a more accurate meta-analysis could have been conducted, such as with hierarchical modeling.

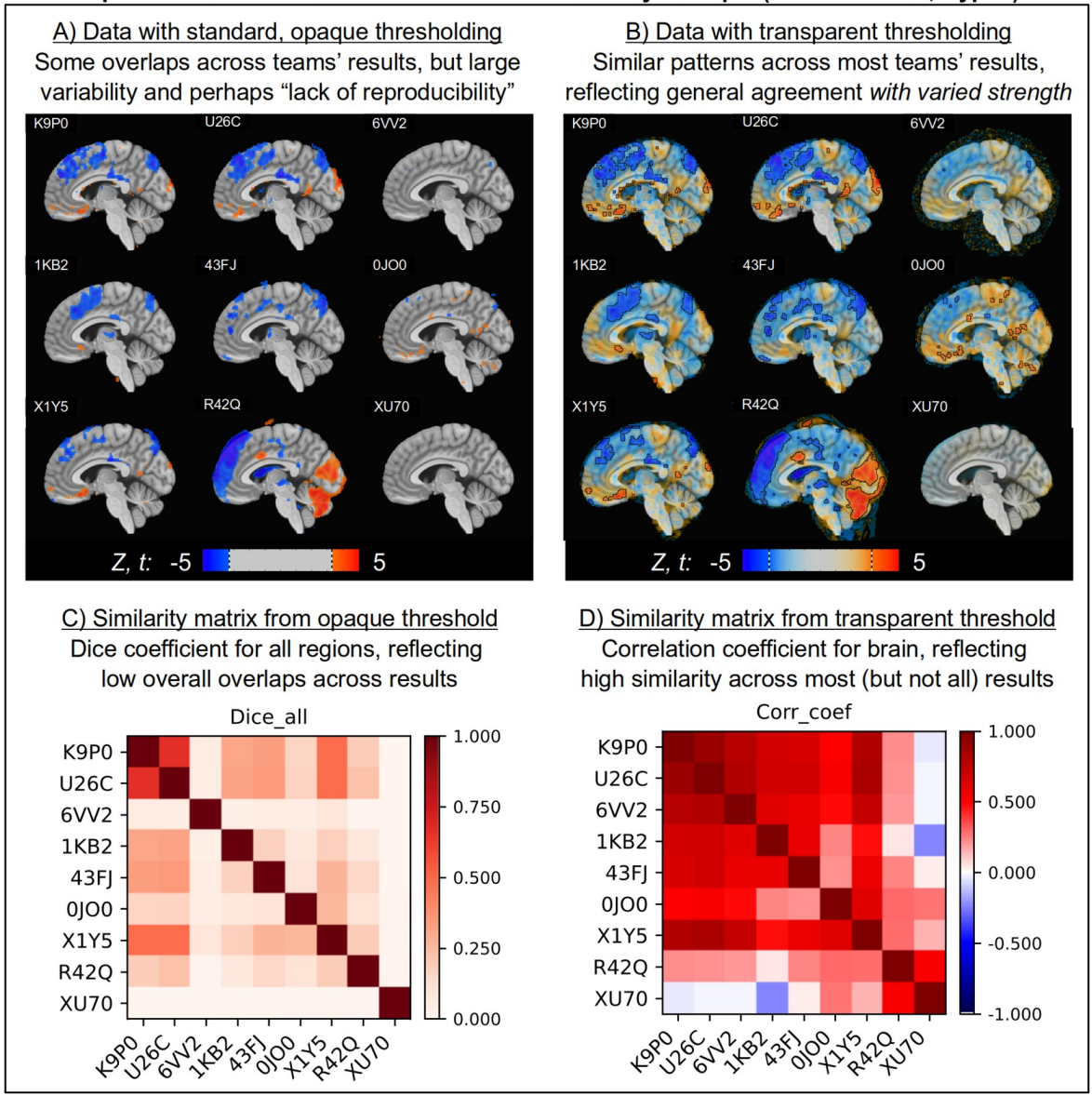

*Figure 2.* Visualizing results from 9 teams who participated in NARPS (Botvinik-Nezer et al., 2020); team IDs are shown in each panel. Panels A-B show Z and t statistic maps in the same sagittal slice of MNI space and thresholded at |Z| or |t| = 3. Panel A shows the results with opaque thresholding (in line with the study's primary meta-analysis), which suggest high variability, inconsistency and disagreement across teams. Panel C shows the corresponding similarity matrix (using Dice coefficients for the binarized cluster maps), which quantifies the generally poor agreement. Panel B shows the same data with transparent thresholding (in line with the study's second meta-analysis), where it becomes apparent that the results actually agree strongly for most subjects, but with varied strength. Panel D shows the corresponding similarity matrix (using Pearson correlation for the continuous statistic maps), showing the typically higher similarity. Transparent thresholding does not uniformly increase similarity, but allows for clearer interpretation of real differences (e.g., bottom right image). Opaque thresholding biases towards dissimilarity (e.g., 3rd column, top and middle). See Taylor et al.

*(2023) for similar comparisons across the full set of NARPS teams and hypotheses, where the same patterns hold.*

*Example 3: context to reduce hypersensitivity and instability of results*

In this example, we contrast the hypersensitivity of opaquely thresholded results to non-physiological features with the relative stability of transparently thresholded ones. Aspects of this have been shown above in the discussion of the NARPS data, but here we demonstrate this point even more directly with a simple case of varying the number of subjects in a study.

Fig. 3 displays the results of a standard one-group analysis with the NARPS data (for Hyp. 2), using a two-sided *t*-test with cluster-based FWE = 5%. Clusters are outlined in white for visibility. The top row contains the results from analyzing the full set of 47 subjects used for group analysis after processing and quality control. Subsequent rows show results if the group size had been reduced by just one subject (arbitrarily chosen by order of subject ID). Fig. 3A shows the results with opaque thresholding applied—both the coverage and number of clusters change notably from row to row. In one case, removing just one subject *decreased* the number of clusters by 25%, and in another case it *increased* the count by 18%. The changes are not simply monotonic or convergent, and one cluster disappears and then reappears (in the left inferior parietal lobule; magenta arrow). A similarity matrix of the clusters (bottom row) shows the amount of variability across an extended set of group sizes (down to 17). Clearly, any interpretation of results with this opaque thresholding will be quite sensitive to group size.

Fig. 3B displays the same results using transparent thresholding. The images are considerably more consistent across the small group size changes, reflecting less sensitivity to the arbitrary differences (e.g., quality control criteria, censoring thresholds, subject motion values, etc.). Note that the changes in cluster count are still known and still vary in the same way, but the contextualized maps allow for the reader's evaluation to remain appropriately consistent and stable. The similarity matrix for these results (Fig. 3D) shows uniformly quite high values even down to the smallest group size. Beyond stability, additional benefits of transparency include having knowledge of the context itself. For example, the area highlighted with the magenta arrow appears to have notable left-right symmetry, which would be unknown in the opaque thresholding case.

Study results are always produced and examined in the context of prior information. The hypersensitivity of opaquely thresholded clusters makes them difficult to rely on for robust comparisons to other papers and even for meaningful evaluation of a study's hypotheses. Opaque thresholding creates the dilemma of determining, for example, the appropriate final number of subjects from which to obtain the "correct" set of clusters. Moreover, it creates a potential incentive for *p*-hacking (Wichert et al., 2016): tweaking and selecting such parameters in a way that might reject the null hypothesis or match more closely with prior work. As shown here, transparent thresholding greatly reduces such temptations, as presented results are more stable and can still be discussed easily, even if just below threshold.

**Reducing sensitivity to arbitrary features (NARPS data, Hyp. 2)**

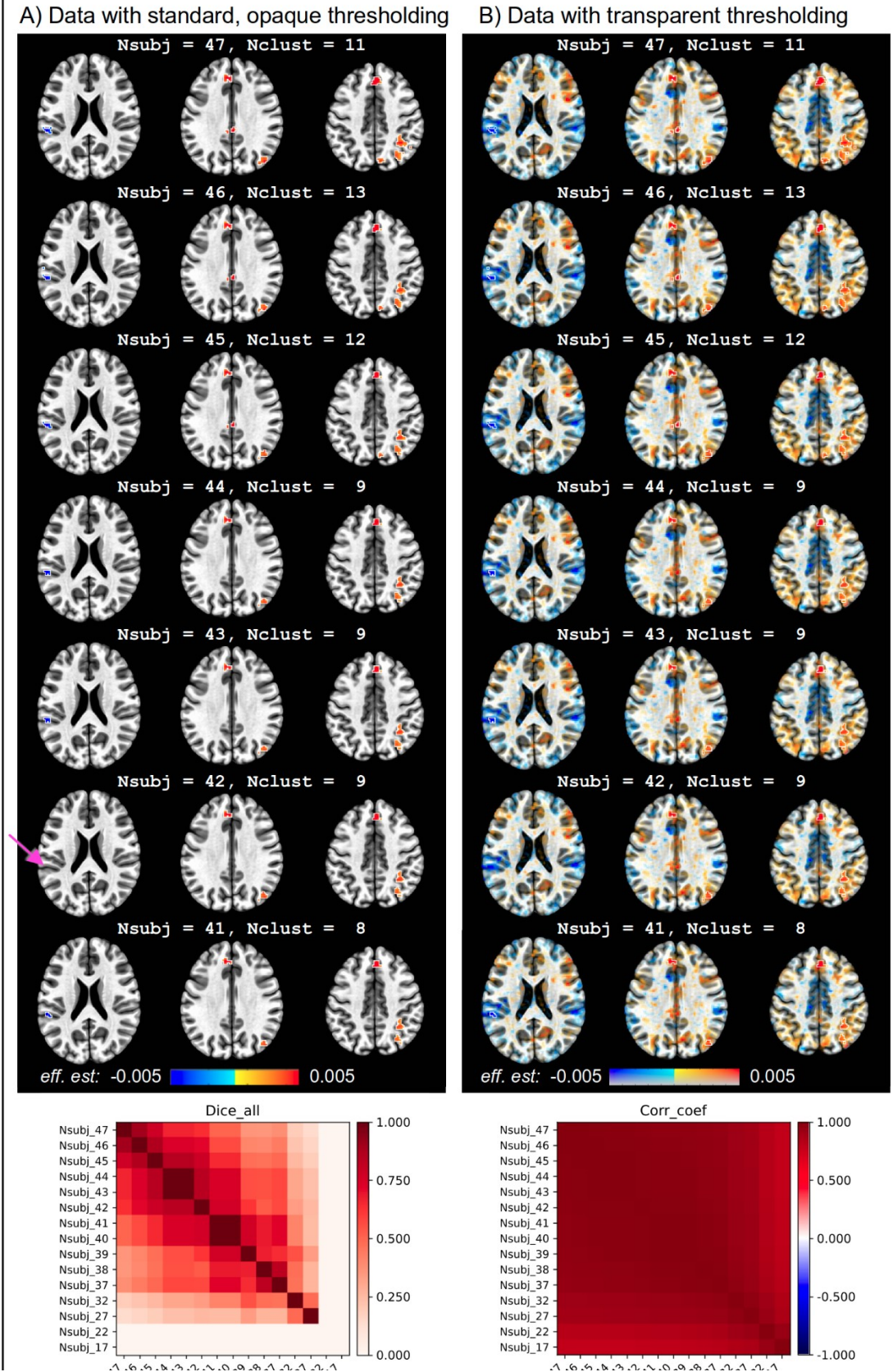

*Figure 3.* *Each panel shows the same axial slices in MNI template space at z = 21S, 32S, 43S (image left = subject left) from NARPS data, Hyp. 2 and 4. The overlay values are effect estimates, in units of BOLD% signal change per dollar for this gambling task, and statistic values were used for thresholding (voxelwise p = 0.001; cluster-level FWE = 5%). All suprathreshold clusters are highlighted with white outlines, for visibility. The top row shows the full group number of subjects (Nsubj), and subsequent rows show results with 1 subject removed. Changes in cluster count (Nclust) are noted for each row. Changes in cluster results —in terms of both coverage and number—are more apparent in Panel A, where opaque thresholding is used. The changes are not simply convergent or monotonic. The magenta arrow highlights a cluster in the left inferior parietal lobule which disappears and reappears with varying Nsubj. The results with transparent thresholding in Panel B are less sensitive to Nsubj changes and also provide useful context. For example, the region highlighted with the magenta arrow appears to have left-right symmetry in negative BOLD response; this information is missed with opaque thresholding. The bottom of each column shows a similarity matrix for each thresholding style (as in Fig. 2), for an extended set of Nsubj. These reflect the striking sensitivity of opaque thresholding (Dice_all, left) with the more stable transparent thresholding (Corr_coef, right).*

*Example 4: using figures and context to avoid misinterpretation*

As a final example of the importance of keeping context in figures to clearly communicate science results, we look back at one of the most widely known studies in FMRI, the "dead salmon study" (Bennett et al., 2009). This study had a single, simple message: when performing massively univariate voxelwise analyses in brain studies, one should adjust for multiple comparisons in some way, such as applying familywise error (FWE) or false discovery rate (FDR) adjustment. This message is clearly stated in the title of the paper, and it is repeatedly restated throughout the abstract and main text. However, the work has still been consistently misreferenced as showing that FMRI is unreliably susceptible to false results (predominantly outside the field) and even frequently misquoted within the field itself.

The part of the paper that unfortunately leaves room for misinterpretation is its lone figure. In practice, figures often leave stronger impressions of results with readers than text. The famous image is reproduced here in Fig. 4A, along with relevant descriptive information summarized from the original caption. The image shows opaquely thresholded results[3] *before* adjusting for multiple comparisons, but not *after* doing so. The authors did perform multiple comparisons adjustment—indeed, that is the analysis step they are promoting—but they simply stated its outcome of "no clusters" *in the text only*. This latter part appears to be ignored by readers relatively frequently, leaving a false impression from the lone, pre-adjustment figure.

One clarifying step would be to include both the "before" and "after" cases in the image, such as in Fig. 4B. The figure's primary message is now more clearly in line with the methodology and the authors' intended purpose.

---

3  This study pre-dates the publication of the transparent thresholding idea by Allen et al. (2012).

But the results reporting could still be improved further by using transparent thresholding. This is shown in Fig. 4D for a different fish,[4] scanned during a standard flashing checkerboard stimulus paradigm (see Supplements for details). We again include both the "before" and "after" cases of statistical adjustment, which match those of the original dead salmon (Fig. 4C shows the new validation salmon results with opaque thresholding). By including subthreshold modeling results in the images in Fig. 4D, it is immediately apparent that the overall pattern is obviously quite noisy and unrelated to structure. This context is useful because it is quite possible that a noisy cluster *could still survive* all of the statistical adjustment and thresholding processes. Seeing the noisy subthreshold context would then provide useful (if not necessary) evidence that any such cluster was likely to be noise-related rather than a function-related cluster (similar to Image 4 in Fig. 1D).

In addition to transparent thresholding, Fig. 4D includes additional useful features. First, It includes data from the whole field of view, so one can judge the pattern of results inside the brain with respect to the "noise floor" background (Taylor et al., 2023). Second, the overlay is the effect estimate dataset (BOLD percent signal change), rather than just the statistics, which are still used for thresholding (Chen et al., 2017). In this way, the reader can appreciate that the effect estimate values are quite low within the subject, even though the statistic values there are relatively high in many places. These features apply even beyond applying transparent thresholds to "before" and "after" images, and should be used whenever possible. For example, seeing the very small effect size of a cluster that happened to survive would provide further useful evidence as to its true, noisy nature; this would apply to the cluster in the lower part of the image in Fig. 4C, if it had been just slightly larger. Thresholding transparently, displaying the effect estimate, and including the background FOV provide useful contextual information to solidify and clarify the interpretation.

---

[4] The dataset from the original dead salmon study has been lost "upstream," but this presents the opportunity to verify that the original findings replicate in another dead fish.

**A) Original dead salmon figure and caption: only results before FWE adjustment**

Thresholding:
voxelwise *p*<0.001 and 3 voxel cluster extent

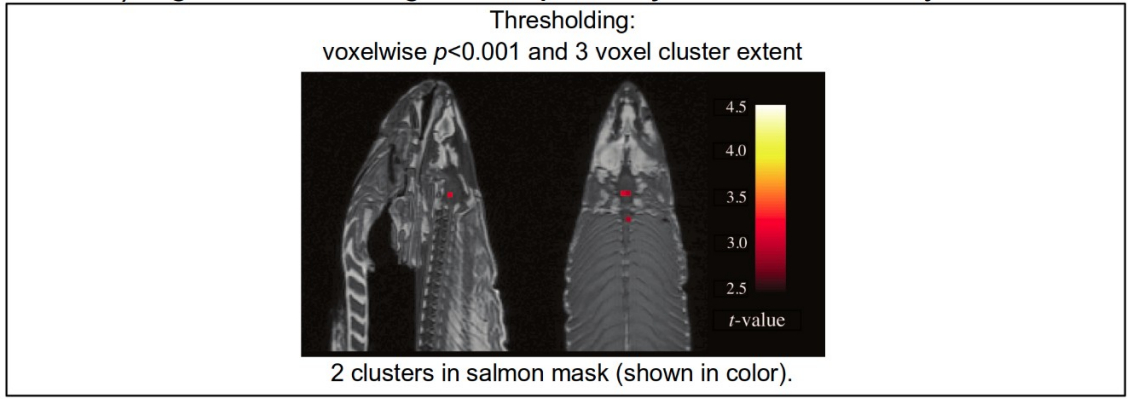

2 clusters in salmon mask (shown in color).

**B) Original dead salmon: also show results after FWE adjustment, for more clarity**

| Unadjusted thresholding: voxelwise *p*<0.001 and 3 voxel cluster extent | Adjusted thresholding (recommended): voxelwise *p*<0.001 and FWE α=0.05 |

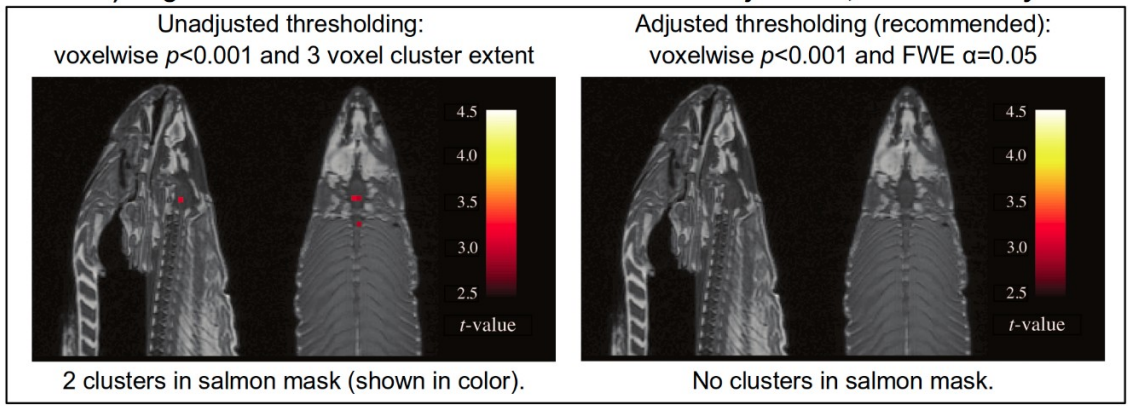

2 clusters in salmon mask (shown in color).  |  No clusters in salmon mask.

**C) New dead salmon: replicate "B" with opaque thresholding before and after FWE adjustment**

| Unadjusted thresholding: voxelwise *p*<0.001 and 3 voxel cluster extent | Adjusted thresholding (recommended): voxelwise *p*<0.001 and FWE α=0.05 |

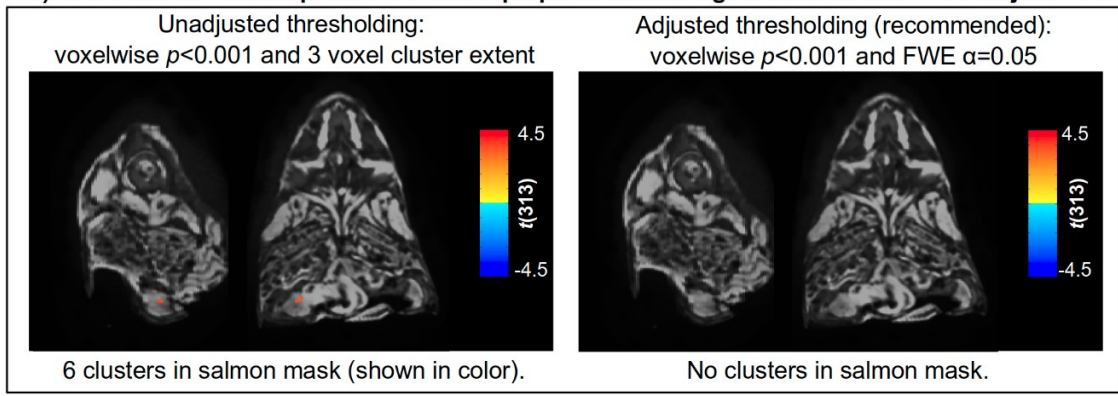

6 clusters in salmon mask (shown in color).  |  No clusters in salmon mask.

**D) New dead salmon: transparent thresholding across whole FOV, with effect estimate overlay**

| Unadjusted thresholding: voxelwise *p*<0.001 and 3 voxel cluster extent | Adjusted thresholding (recommended): voxelwise *p*<0.001 and FWE α=0.05 |

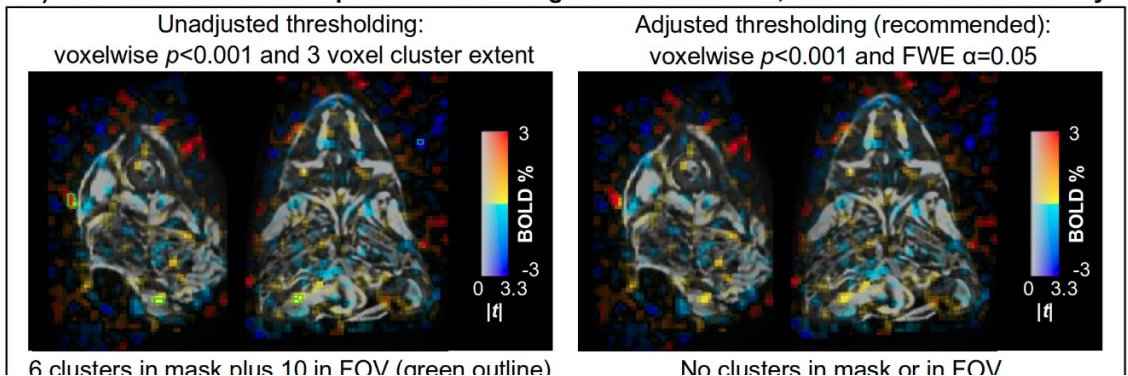

6 clusters in mask plus 10 in FOV (green outline)  |  No clusters in mask or in FOV

***Figure 4.** Panel A shows the lone figure from the famous "dead salmon study" (Bennett et al., 2009; with permission of the authors). The figure is opaquely thresholded and only shows results* before *the recommended multiple comparisons adjustment; by not including the* after *image, many readers have misinterpreted the overall study message, even though it is clearly repeated throughout the text. Panel B shows a simple improvement to make the figure's message clearer and reduce the likelihood of misinterpretation, by including both before- and after-adjustment images. Panel C shows the new validation salmon in the same manner as Panel B, replicating the original results with opaque thresholding. Panel D shows how more complete context can be added to further reduce risks of misinterpretation by thresholding transparently (suprathreshold regions outlined in green), displaying the effect estimate in units of BOLD % signal change as overlay colors, and even showing results outside the subject anatomy. This extra information would provide valuable evidence that any cluster that might survive here—which is possible even when including multiple comparisons adjustment—is likely noise due to the background pattern, high noise floor and likely low effect estimate value.*

**DISCUSSION**

Neuroimaging authors need to supply enough details for the analysis to be understood and replicable (e.g., Maumet et al., 2016; Nichols et al., 2017). They also need to balance presenting "digestible" results with retaining meaningful information content (Chen, Taylor, et al. 2022). The question of how much data to present and in what form is important, and we can turn again to Anscombe (1973), who commented on the value of graphs to provide useful contextual information beyond just summary statistics:

> *Graphs can have various purposes, such as: (i) to help us perceive and appreciate some broad features of the data, (ii) to let us look behind those broad features and see what else is there. Most kinds of statistical calculation rest on assumptions about the behavior of the data. Those assumptions may be false, and then the calculations may be misleading. We ought always to try to check whether the assumptions are reasonably correct; and if they are wrong we ought to be able to perceive in what ways they are wrong. Graphs are very valuable for these purposes.*

In a near-exact parallel, we believe that transparent thresholding provides important value for understanding and evaluating results in neuroimaging. Many assumptions associated with opaque thresholding are inconsistent with the data, thereby increasing odds of misinterpretation and biases. In contrast, retaining context provides the benefits of both "appreciating broad features of the data" and letting readers "see what else is there."

The four examples presented here illustrate these points and other benefits of retaining context in neuroimaging results. The primary benefit is to provide a clearer, deeper, and more accurate understanding of the data, for both authors and readers. This is the ultimate goal of a scientific experiment. Results reporting should prioritize having comprehensive information over artificial dichotomization. Removing context with opaque thresholding inserts a large amount of ambiguity and bias into results, tilting the scales towards misinterpretation.

The benefits of transparent thresholding—and costs of opaque thresholding—apply across modality and species. Whether analyzing FMRI, diffusion weighted imaging (DWI) or PET (Bishay et al., 2024) data in humans, macaques (Russ and Leopold, 2015) or rodents, showing subthreshold context improves interpretability. It can also be applied equivalently to voxelwise or ROI-based analyses (Chen et al., 2020; Taylor et al., 2023), as well as to both volumetric and surface-based analyses (Sava-Segal et al., 2024; Freund et al., 2025).

The issue of thresholding images touches at the core of the scientific endeavor, particularly for neuroimaging. An individual study rarely provides a definitive answer. Instead, empirical science is an iterative process that builds upon cumulative evidence from multiple studies. Using opaque thresholds treats an individual study as a standalone decision-making tool, which misrepresents the essence of scientific inquiry. In contrast, transparent thresholding assists that process by presenting results as widely informative evidence rather than as a narrowly defined "answer". It is also directly in line with larger mathematical recommendations about improving the use and interpretation of statistics and $p$-values across broad scientific disciplines, which include having an "interpretation of results in context" and making "complete reporting" (Wasserstein and Lazar, 2016). By focusing on the broader context and appropriately including results that simply have higher uncertainty, researchers can better align with the collaborative and progressive nature of empirical investigation.

As demonstrated below in Figs 5-7, there are now a large number of software packages that make a similar form of transparent thresholding available to researchers. These encompass implementations for volumetric, surface- and region-based studies. Several of these were added or streamlined as part of this work. This methodological accessibility is an important practical step for the neuroimaging field (and it will likely continue to grow), allowing researchers to easily adopt the same form of transparent thresholding for their image visualization.

This approach is a useful complement to data sharing (such as via Neurovault (Gorgolewski et al., 2015), OSF, or another resource), which itself is beneficial to the field, but applying transparent thresholding is distinct and important on its own. Figures have a powerful and primary role in interpreting results, and therefore they should be as informative as possible within the presented publication in order to facilitate accurate evaluation. The examples presented above have demonstrated this. Showing opaque figures-as-usual and relying on readers to download and visualize the data again separately does not accomplish this effectively.

*Reducing biases*

Standard opaque thresholding, while done with well-meaning intentions, often inserts bias into subsequent analysis and meta-analysis. This practice both harms the ability to accurately assess reproducibility and acts to decrease reproducibility unnecessarily. The analyses of the NARPS data show both of these features, as directly demonstrated in Taylor et al. (2023) and the examples above, as well as indirectly noted in Botvinik-Nezer et al. (2020). As shown above, transparent thresholding showed increased reproducibility where appropriate (i.e., when results agreed but at different strengths) and showed low reproducibility where appropriate (when results disagreed or were essentially null). Thresholding is a processing choice, and these studies together strongly suggest that opaque thresholding is often detrimental to analyses and meta-analyses. Reproducibility has been a long-discussed topic in the field, and retaining context in results and figures is a clear step to help address it.

The adoption of transparent thresholding helps reduce several other biases in neuroimaging reporting, while still preserving the ability to highlight the most significant regions:
- Type II errors, which may arise from overly strict multiple comparisons adjustments (Cremers et al., 2017), are reduced because subthreshold regions (particularly those just below the cutoff) are still visible. Such regions can still be assessed in the context of other studies or prior knowledge, without simply being treated (inaccurately) as "no effect".
- As discussed above, transparent thresholding greatly reduces incentives for *p*-hacking (Wicherts et al., 2016) to be able to report results in predetermined regions, or relatedly (and problematically) to "spin" results (Boutron and Ravaud, 2018).
- Publication bias or the "file drawer problem" exists in neuroimaging, where findings with weak or limited statistical evidence are typically not reported. The result is that many meta-analyses misrepresent assessments (Jennings and Van Horn, 2012; Ioannidis et al., 2014). Transparent thresholding offers a way that findings with weak statistical evidence can be assessed and reported more informatively. That is, having multiple related studies with visible sub-threshold responses, represents critical information to include in meta-analyses. Additionally, displaying subthreshold regions meaningfully, even when some locations have suprathreshold ones, can help reduce publication bias.
- Statistical thresholding itself is essentially a form of selective reporting, which was selected as the top factor contributing to reproducibility problems according to a recent Nature journal "reproducibility survey" of researchers (NPG, 2018). Consequently, overly stringent statistical thresholds and opaque thresholding may hinder rather than help the scientific process, as they can obscure meaningful patterns and impede the synthesis of findings across studies.

In addition to reducing those biases, the improved stability of transparent thresholding (see Ex. 3, above) greatly benefits studies of difficult to scan populations. While there has been a movement to increase group sizes, animal imaging studies (e.g., of nonhuman primates and rodents) remain small; in many cases, these have less than 10 subjects (Mandino et al., 2020), though these often have multiple sessions. In human clinical studies, group sizes also tend to be much smaller than standard research studies (Szucs and Ioannidis, 2020). The accrual of clinical participants is often limited by practical considerations of availability (e.g., rare diseases), cost (travel), complexity (health considerations and medical monitoring), and study length (drug trials). In these cases, opaque thresholding practices often result in few (or even no) regions above the statistical cut-off, potentially resulting in an inability to publish (i.e., publication bias) or an incentive to *p*-hack. Thus, the results that can be reported with typical thresholding are generally limited and tightly bound to the conditions of the sample itself, making them difficult to interpret and reproduce. (Data sharing is also typically more challenging for clinical studies.) While increased stability of results is not a substitute for having an adequately powered sample, transparent thresholding enables these smaller studies, acquired under difficult conditions, to contribute to the literature in a more meaningful way rather than adding to the "file drawer problem." With transparent thresholding, the regions with larger uncertainty are clearly viewed as such, rather than being hidden or artificially pushed above thresholding by statistical maneuvering. At the same time, transparent thresholding helps reduce the likelihood of false negatives, which can be particularly important in clinical studies. It also parallels the trend in the clinical literature of moving away from the thresholding dichotomy, such as viewing results at multiple thresholds instead of a single arbitrary threshold (Voets, et al., 2025).

*Forward looking goals*

As another important focus on figures, we note that papers are increasingly extracted and processed by algorithms. NeuroSynth (Yarkoni et al., 2011) was an early example of a tool that automatically parsed publication text and aggregated information together. Today, large language models (LLMs) and multimodal LLMs (MLLMs) are increasingly applied to databases and libraries—some even have a specific focus on medical imaging—creating summary tools from both text and figures (Bhayana, 2024; Bzdok et al., 2024). Just as retaining accurate descriptions in text improves the results of these tools, so does (or surely will) having more informative figures.

The adoption of transparent thresholding facilitates and complements meta-analyses that include more data in neuroimaging. For example, auxiliary scatterplots of statistics and effects in MVPA results have revealed interesting spatial patterns (see Fig 4 of Visconti di Oleggio Castello et al., (2017)). If whole brain results are shared and used for cross-study comparisons, it is more consistent to have the results shown across the whole brain in the first place. It would be confusing and inconsistent to see meta-analyses point out differences in regions that were hidden in the initial papers. Moreover, proposed multiverse approaches (e.g., Dafflon et al., 2022; Lefort-Besnard et al., 2024) aim to combine results from multiple processing pipelines or statistical methods for a given dataset, similar to meta-analyses. Transparent thresholding facilitates reporting such "doubly probabilistic" results across the whole brain, where one would expect different sub-tests to have meaningful evidence to be reported.

Another way to improve both cross-study meta-analyses and within-study interpretations is to include effect estimates in the results, rather than only showing statistics (Halsey et al., 2015). Many neuroimaging modalities have physical units, such as DWI, and some, like FMRI, can be scaled to have meaningful units (Chen et al., 2017; Flournoy et al., 2020). These provide separate information about the data and its modeling, including the *practical* significance of results. In most areas of science, it would be inconceivable not to include effect estimates, as they form the basis of analysis and interpretation. Leaving these measures out wastes information (Chen et al., 2022) and increases the ambiguity of results. For example, showing effect estimates provides useful evidence about whether suprathreshold locations are more likely true or false positives (as noted in Ex. 4). They also enable more meaningful comparisons in meta-analyses than ones based on statistics alone (e.g., Maumet and Nichols, 2016).

Finally, transparent thresholding directly benefits quality control (QC) efforts during both data processing and results presentation. Several examples of this were provided in Reynolds et al. (2023), where artifacts in the acquired EPI time series would likely have gone unnoticed without using transparent thresholding in the creation of seed-based correlation QC images. In both Ex. 1 and 4 here, subthreshold patterns helped distinguish when any results might likely be due to noise or artifact, reducing the risk of false positives. In these and other cases, wider results reporting and retention of context provide greater confidence and clearer interpretability of figures.

*Addressing comments, concerns and questions about transparent thresholding*

While some in the neuroimaging field have been enthusiastic about using transparent thresholding to present more informative and less ambiguous results, others have been skeptical or raised critiques. Here we summarize some of the latter and address these points.

**1) Thresholding at exactly p=0.001 and FWE = 5% is rigorous, and reporting any other results will harm reproducibility with false positives.** Firstly, the proposed transparent thresholding still highlights the exact same regions above a given threshold (with full opacity and outlining). Secondly, threshold values themselves are typically set by convention, with various round numbers argued for at various points in history. Fisher initially wrote about $p = 0.05$ as useful, primarily because it corresponds to a round, two-tailed test value $Z \approx 2$, but he also used other values such as $p = 0.01$ (Fisher, 1925). Many of his contemporary statisticians viewed such threshold choices as arbitrary and obfuscating (see Kennedy-Shaffer, 2019). More recently, one group of statisticians pushed to lower the canonical $p$-threshold to a still different value 0.005 (Benjamin et al., 2018), while another proposed rejecting $p$-value thresholds altogether (Amrhein & McShane, 2019). Even for clinical FMRI, there is no standard thresholding practice (Voets et al. 2025). This snapshot of a hundred-year-old debate alone shows there is no single, canonical threshold between "significance" and "insignificance," and many modern statisticians view $p$-values as an unreliable focus (Halsey et al., 2015). Entirely hiding a cluster that has FWE= 5.01% is arbitrary and unscientific—such a practice itself actually harms reproducibility in the long term.

**2) A given threshold may be arbitrary, but if everyone uses the same value, then results will still be on equal footing for comparisons.** FMRI and many other kinds of neuroimaging data are noisy, with noise profiles changing across the field of view and with the scanner used. The underlying biological responses are continuous with varying magnitudes. In practice, no threshold will be consistent across studies, e.g. due to differing sample size and power, or noise characteristics and variability, or evolving acquisition strategies. Instead, applying a strict threshold for a continuous response variable also greatly increases sensitivity to non-physiological differences across studies, such as scanner type, field strength, number of subjects (see Ex. 3 above), voxel size, trial number/length, field inhomogeneities, modeling methods, etc. Opaque thresholding will strongly bias comparisons and meta-analyses towards irreproducibility and non-replicability (see the NARPS data discussion, above).

**3) Science is about "storytelling" and transparent thresholding complicates the story of results by showing more things.** Few stories of note have only main actors and no supporting cast and context—*Romeo and Juliet* (Shakespeare, 1597) would be a poor play if it contained no other characters. Perhaps fully *un*thresholded results are overly complicated to interpret for most, but transparently thresholded ones primarily add just near-significant regions and larger context. If there are not many near-threshold results, then the story stays terse. Alternatively, if there *are* many near-threshold results, then that is *part* of the story. In either scenario the reader learns from seeing the additional context. It will both facilitate storytelling and clarify a narrative, such as by suggesting which network a cluster belongs to. In Ex. 1, the additional context from transparent thresholding presents compelling evidence to challenge the narrative of strong laterality that opaque thresholding would have produced. As Einstein (probably) noted about science, "Everything should be made as simple as possible, but not simpler" (Shapiro, 2006). Storytelling should not be used as an excuse to sacrifice meaningful information.

**4) We applied transparent thresholding, and now see too many regions to discuss—the paper will be too long.** It does not seem necessary to have a detailed discussion about every single region with visible, subthreshold results. In many papers, researchers do not even write about each suprathreshold region. It would be logical to highlight any regions of particular interest, as determined from either prior research or background knowledge, and then the rest can remain as observable context and/or for potential relevance to future studies. Transparent thresholding also allows for further discussion of regions that show *no* visible response with transparent thresholding, especially if activity had been hypothesized there. In total, this should not make discussions more burdensome but instead more informative and clearer. It should also facilitate connecting the results to those of existing literature, functional networks and prior domain knowledge, by providing more globally informative results. (Additionally, if authors are concerned that a figure will not get published with transparent thresholding because there is a mess of blobs and artifacts near-threshold, this is a problem with their data and not something to be swept under the rug with opaque thresholding. Hiding issues in data to facilitate publication is generally considered poor scientific practice.)

**5) We checked out the results at multiple thresholds within our group, so we feel confident about publishing with standard thresholding.** Readers engage with scientific articles critically, a procedure that is facilitated by seeing more of the underlying evidence for themselves as they read. If transparent thresholding reveals no near-threshold results, then that simply reinforces the authors' interpretation—nothing is lost. If transparency reveals some additional locations of interest, then readers will be aware even if it does not figure strongly into the authors' interpretation. In fact, transparent thresholding should be viewed as a *helpful* tool to convince readers. For example, if the authors have used multiple thresholds and confirmed for themselves that their suprathreshold region is not just an extension of a near-threshold blob from the CSF, then they only strengthen their argument by including this information in their figures. In science, it benefits both the authors and readers to present the full picture. Moreover, as shown in Ex. 3, transparent thresholding will reduce the hypersensitivity of near-threshold

results to arbitrary parameters (like number of subjects) and the incentives for *p*-hacking around varied thresholds.

**6) We will upload the unthresholded results to a public repository, so we will publish opaque images and people can explore the data themselves later.** Making full results public is great[5] and certainly enables meta-analyses, but separating the more complete results from the paper greatly diminishes the ability for critical understanding by the reader. It also places a burden of time and effort that not every reader will go through.[6] A study should provide strong evidence for its interpretations, and transparent thresholding does a better job of this than opaque thresholding. Two well-known aphorisms apply:
- *A picture is worth a thousand words.* Figures are likely the most important messengers in a scientific study. We provided multiple examples here where there have been critical misinterpretations of written results, simply because figures were either lacking important information or were themselves missing. A summary figure can end up in talks and general discourse in ways that separate it from the complete story of the actual data in the repository and can even distort the intended message of the authors.
- *First impressions are the most important.* Reanalysis of public data might lead to a different interpretation from an initial study, but there will be a long lag before that update can enter the scientific conversation. Consider the widely cited NARPS study, which created a strong impression of poor FMRI reproducibility that is still echoed today. The follow-up analysis of the public data by Taylor et al. (2023) showing strong evidence for a different message was published three years later, in a separate journal, and with much less impact.

**7) I'm a clinician. I need to know definite regions.** You are the expert for your work, so *you* should be the decision maker from a reasonable set of evidence. Transparent thresholding presents meaningful results for you to interpret and from which to make informed judgments. In contrast, starting with opaque thresholding presents a predetermined decision based on an arbitrary cut-off and removes potentially useful information from your consideration—in short, it puts the scalpel in the hands of an academic. It is more scientific (and likely better for clinical outcomes) to share contextualized evidence for clinicians to interpret for their purposes. Some diagnostic specialties even use fully *un*thresholded maps regularly, but when greater digestibility is required, transparent thresholding helps reduce the risk of false negatives and misinterpretation. Consider the divergent laterality findings in Ex. 1. In special cases that a binarized image is needed (e.g., in surgery), then that can still be derived from a transparently thresholded one and likely with more confidence about the localization. Voets et al. (2025) discuss further issues for clinical applications.

---

5 When possible—not all datasets can be shared, particularly in clinical studies.
6 An earlier version of this draft purposely omitted images of the original Anscombe Quartet for readers to look up online, as a simplified example of the burden caused by separating relevant figures from the main text. However, even this simple search (no downloads or software needed) was deemed too frustrating by coauthors, so the images were included here.

**8) I write for a non-technical audience, and I need to show a direct story for those outside the field.** Even for non-neuroimaging readers, oversimplifying the results is problematic. Experience with the dead salmon (Ex. 4, above) and other studies has shown that. That manuscript made a clear and valid point, but sharing only the one opaquely thresholded image has resulted in repeated misinterpretations of their core message. With transparent thresholding and the other features discussed above, we posit that the dead salmon study would have been less easy to misinterpret. While its coverage in non-technical media might not have become as expansive, those it did reach would have gained a better understanding of science and the core issues. (And it would still have received wide attention within the field, because it is a great demonstration of a serious issue.) Scientific understanding should be as accurate as possible, for both those in the field and those outside of it.

**9) Reviewers have criticized showing subthreshold results, complained about seeing more complicated spatial patterns, and do not like the lines around regions—this makes me hesitant to try to publish with this approach.** While many researchers have successfully adopted transparent thresholding in the publications (see a partial list in Table S1), it does take time for new ideas to become accepted and commonplace. We hope that articles like this, which show transparent thresholding's many benefits and demonstrate the significant problems with opaque thresholding, will help this way of displaying results become more commonplace. If spatial patterns are complicated because many results are slightly subthreshold, then it is likely even more important to use transparency, to reduce the hypersensitivity to non-physiological features and chance of misinterpretation, as shown above in Ex. 3. While transparent thresholding is not currently normative, it is plausible that manuscripts with opaque thresholding should or will eventually themselves be critiqued over what is *not* shown to readers. We hope that the examples and points raised in this paper, as well as those in Allen et al. (2012), Chen et al. (2022), Taylor et al. (2023), and Sundermann et al. (2024), can provide convincing rationales to the reviewers for this approach.

**10) It is too difficult to implement this visualization.** Transparent thresholding in data visualization is now available in a wide number of publicly available neuroimaging software packages (as well as in separately programmed implementations): in the original Trends-Matlab toolbox (https://trendscenter.org/x/datavis; Allen et al., 2012) and the related GIFT (http://trendscenter.org/software/gift); in AFNI (Cox, 1996) and preliminarily in the surface-based visualization of SUMA (Saad et al., 2004; Saad and Reynolds, 2012); in BrainVoyager (Goebel, 2012); in FSL's FSLeyes (McCarthy, 2024; Smith et al., 2004); in NiiVue (Hanayik et al., 2023); in RMINC (Lerch et al., 2017) and the related MRIcrotome (https://github.com/Mouse-Imaging-Centre/MRIcrotome); in CIVET (Ad-Dab'bagh et al., 2006) and the related minc-toolkit-v2 (https://github.com/BIC-MNI/minc-toolkit-v2); in Nilearn (Nilearn contributors, 2025); and in bidspm (https://github.com/cpp-lln-lab/bidspm). Representative images are shown in Figs. 5-7, which encompass volumetric, surface-based and ROI-based cases. See Table S1 in the Supplements for further examples.[7] We hope that increased use of transparent thresholding will see its implementations spread further.

---

7 Interestingly, the majority of these studies that used transparent thresholding also presented effect estimates as overlays, providing further useful information as recommended here (see Ex. 4).

**Examples of transparent thresholding in available software implementations**

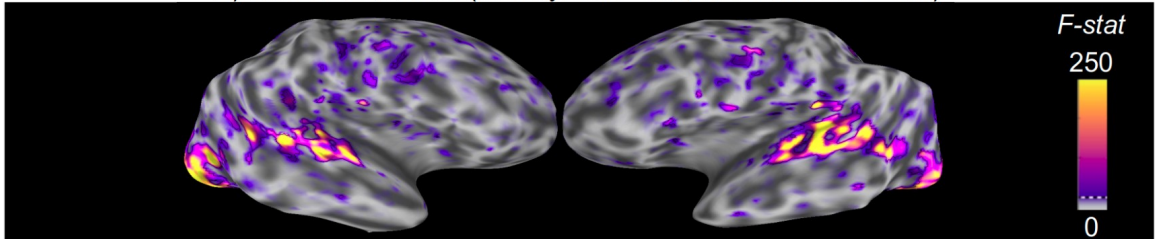

A) SUMA: task FMRI (overlay: full *F*-stat, threshold: full *F*-stat)

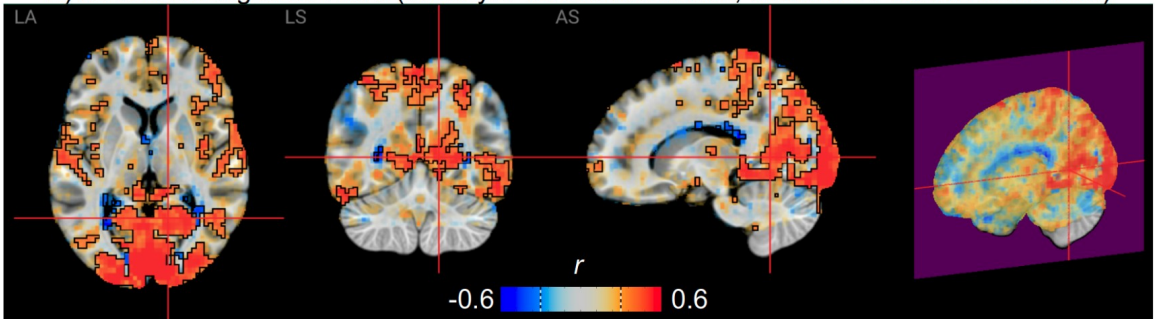

B) NiiVue: resting state FMRI (overlay: Pearson correlation, threshold: Pearson correlation)

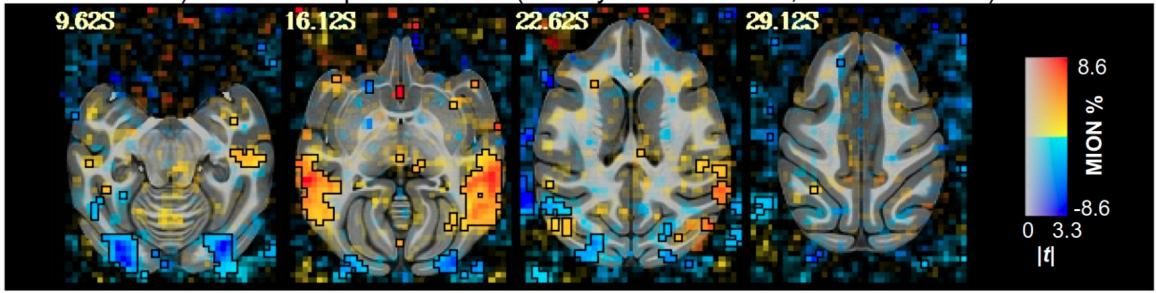

C) AFNI: macaque task FMRI (overlay: effect contrast, threshold: *t*-stat)

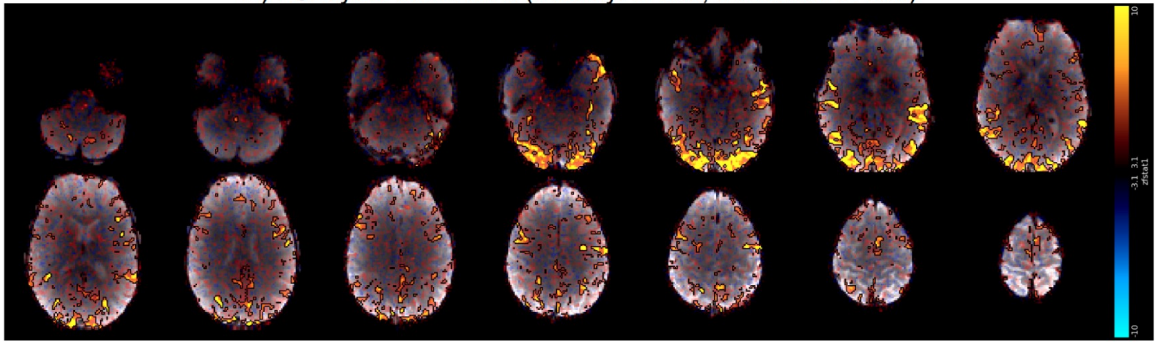

D) FSLeyes: task FMRI (overlay: *Z*-stat, threshold: *Z*-stat)

*Figure 5.* Example images of transparent thresholding from various software implementations (and see Figs. 6 and 7 for more examples). Descriptions of the data and software usage are provided in the Supplements.

**Examples of transparent thresholding in available software implementations**

A) Trends-Matlab & GIFT: task FMRI (overlay: effect contrast, threshold: *t*-stat)

B) BrainVoyager: task FMRI (overlay: *t*-stat, threshold: FDR *q*-values)

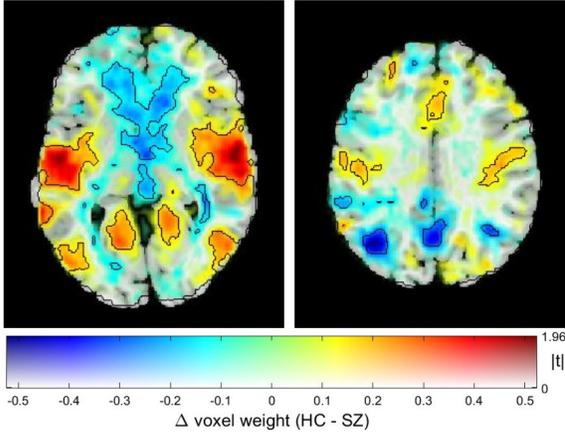
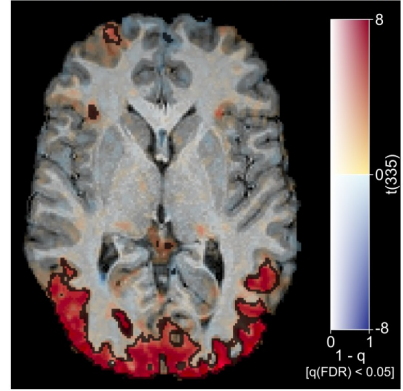

C) CIVET & minc-toolkit-v2: cortical thickness of structural MRI (overlay: *t*-stat, threshold: *t*-stat)

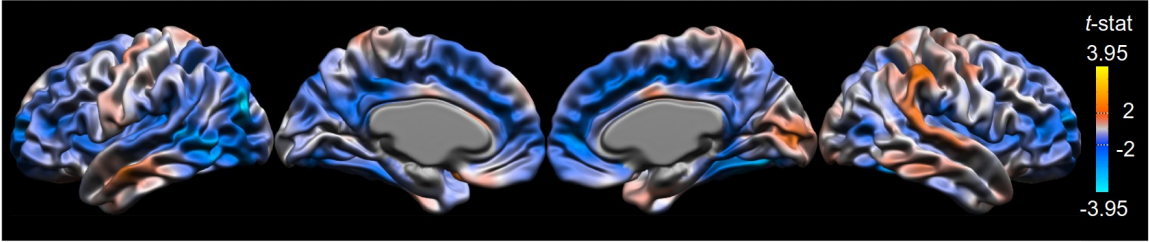

D) RMINC & MRIcrotome: rat structural MRI morphometry (overlay: *t*-stat, threshold: *t*-stat)

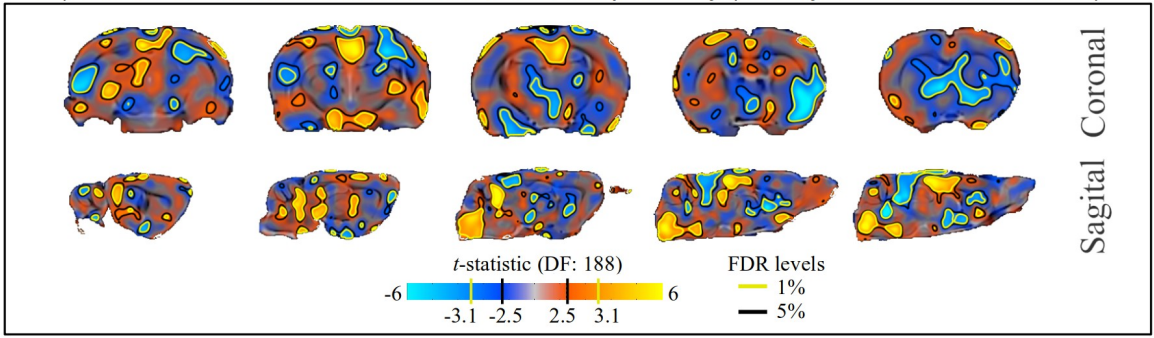

*Figure 6.* Example images of transparent thresholding from various software implementations (and see Figs. 5 and 7 for more examples). Descriptions of the data and software usage are provided in the Supplements.

**Examples of transparent thresholding in available software implementations**

A) RMINC: mouse structural MRI morphometry (overlay: *t*-stat, threshold: *t*-stat)

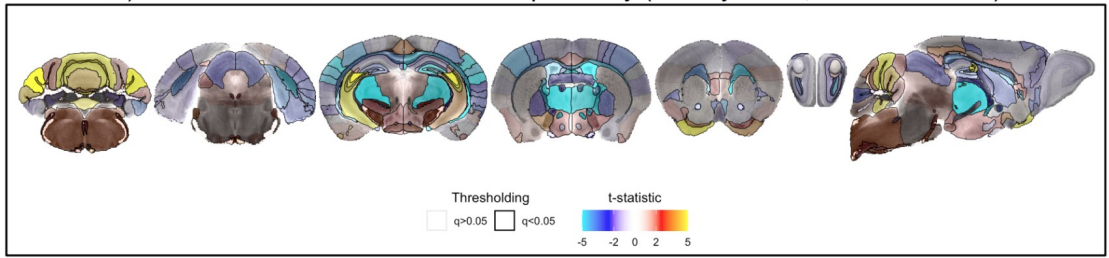

B) AFNI: task FMRI (overlay: effect estimate, threshold: Bayesian statistical evidence)

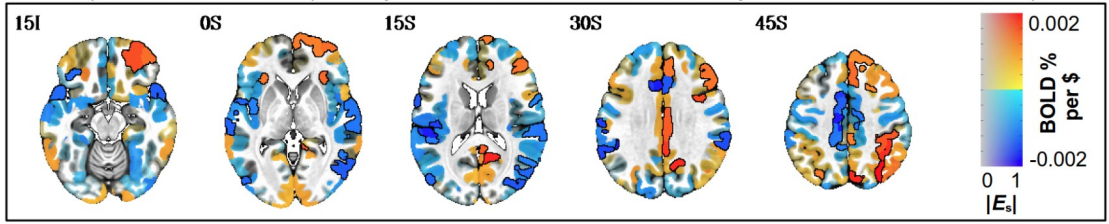

C) bidspm: task FMRI (overlay: *t*-stat, threshold: *t*-stat)

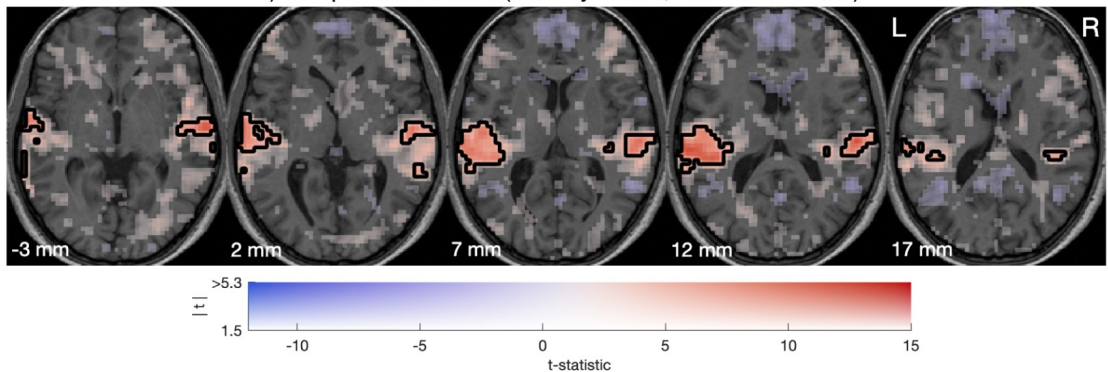

D) Nilearn: task FMRI (overlay: *t*-stat, threshold: *t*-stat)

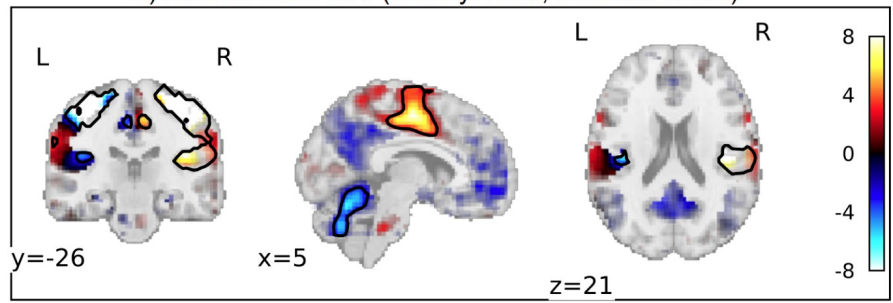

*Figure 7. Example images of transparent thresholding from various software implementations (and see Figs. 5 and 6 for more examples). Descriptions of the data and software usage are provided in the Supplements.*

## CONCLUSION

Choosing how to threshold results is an important processing choice in FMRI and more widely across neuroimaging, including in many clinical applications. Transparent thresholding highlights the same strong regions as standard opaque thresholding but also retains brainwide context that is important for both authors and readers to see. This approach enhances understanding and helps to avoid misinterpretation in figures, which are key to presenting study results. It also provides a better framework for accurately comparing datasets and evaluating reproducibility. In contrast, opaque thresholding introduces strong biases and hypersensitivity to non-physiological features, harming within-study evaluation and cross-study reproducibility. Transparent thresholding is straightforward and easily implementable. In fact, it has already been included in a large number of packages and available scripts, providing an accessible and common way for neuroimagers display data. We hope that researchers in the field will move toward adopting this strategy when presenting results.

**ACKNOWLEDGMENTS**

The present work was completed as part of several authors' official duties as Government employees. The views expressed do not necessarily reflect the views of the National Institutes of Health (NIH), the Department of Health and Human Services (HHS), or the United States Government. We appreciate the authors of the original "dead salmon" study (Bennett et al., 2009) giving their permission for reproduction of the study's figure in this text. We also appreciate the authors of the original NARPS project (Botvinik-Nezer et al., 2020) for making public the results that had been submitted by all participating teams; this has been an important dataset for the neuroimaging community to consider and keep reanalyzing for new perspectives. PAT, DRG, PDL, JKR, RCR, and GC were supported by the NIMH Intramural Research Program (ZICMH002888) of the NIH/HHS, USA. AM was supported by a grant from NIH/NIA (R01AG083919). BER was supported by grants from NIMH (R01MH111439, RF1MH117040, R01MH124045, and P50MH109429) and NINDS (R01NS109498). BT et HA benefited from state aid managed by the Agence Nationale de la Recherche under the France 2030 program (reference ANR-22-PESN-0012), and have also received funding from the European Union's Horizon 2020 Framework Programme for Research and Innovation under the Specific Grant Agreement HORIZON-INFRA-2022-SERV-B-01. CCG was supported by the Basque Government BERC 2022-2025 program, the Spanish State Research Agency through BCBL Severo Ochoa excellence accreditation CEX2020-001010/AEI/10.13039/501100011033, and the project PID2023-149410OB-100 funded by MCIU/AEI/10.13039/501100011033/FEDER, EU. CM has funding from consulting and lecturing for Siemens Healthineers on behalf of Evangelisches Krankenhaus Oldenburg. CP is funded by the Novo Nordisk Foundation (grant NNF20OC0063277). CR was supported by grants from the NIBIB/NIMH (RF1-MH133701) and NIDCD (P50-DC014664). DAH and JGC were supported by NIMH Intramural Research Program (ZIAMH002783); VR was supported by the NIMH Intramural Research Program (ZICMH002884); and PAB was supported by both of these grants. DAL was supported by NIMH Intramural Research Program (1ZICMH002899). EAGV was supported by UNAM PAPIIT projects IN213924 and IA201622, and is part of the CONAHCYT (3256252/629578) and DGAPA (781759) postdoctoral projects. GAD was supported by the Douglas Research Centre and funds from McGill University's Healthy Brains for Healthy Lives initiative (a Canada First Research Excellence Fund Initiative). JWE was supported by NIMH Intramural Research Program (ZIAMH002857). JPL was supported by The Wellcome Centre for Integrative Neuroimaging, which is supported by core funding from the Wellcome Trust (203139/Z/16/Z and 203139/A/16/Z). LDR was supported by the NIH's Office of Intramural Training and Education (OITE) through an Intramural Research Training Award (IRTA), and was also funded by the National Institute of Biomedical and Bioengineering Intramural program: Processing and Analysis of Quantitative Diffusion MRI Data, ZIA EB000088. LP was supported by a grant from NIMH (R01MH071589). MB was supported by a Flagship ERA-NET grant SoundSight (FRS-FNRS PINT-MULTI R.8008.19). MC was supported by the Canadian Institutes for Health Research, National Sciences Research Council of Canada, and HBHL, while also receiving salary support from Fonds de recherche du Québec - Santé. MGB was supported by grants from NIH NICHD (R03HD113915, R21HD108587). OFG and RG have financial interests tied to Brain Innovation company. SM has received funding from the European Union's Horizon 2020 research and innovation programme under the Marie Skłodowska-Curie grant agreement No 101109770. ST was supported by UCSF RAP REAC Award 7504825. TEN was supported by NIH grants (R01DA048993, R01MH096906, R01EB026859, U19AG073585) of the NIH/HHS, USA. TH's primary contributions were made while employed by FMRIB/Oxford University, and were supported by the Wellcome Centre for Integrative Neuroimaging (Oxford, UK) via funding from the Wellcome Trust (203139/Z/16/Z and 203139/A/16/Z) and the NIHR Oxford Health Biomedical Research Centre (NIHR203316); the views expressed are those of the author(s) and not necessarily those of the NIHR or the Department of Health and Social Care. VDC was supported by grants from NIH (R01MH123610) and NSF (2112455). YH was supported by a grant from NIH-NIBIB (P41EB019936). This work utilized the computational resources of the NIH HPC Biowulf cluster (https://hpc.nih.gov).

**REFERENCES**

Ad-Dab'bagh Y, Einarson D, Lyttelton O, Muehlboeck J-S, Mok K, Ivanov O, Vincent RD, Lepage C, Lerch J, Fombonne E, Evans AC (2006). The CIVET image-processing environment: A fully automated comprehensive pipeline for anatomical neuroimaging research. Proc. OHBM-2006. http://www.bic.mni.mcgill.ca/users/yaddab/Yasser-HBM2006-Poster.pdf

Allen EA, Erhardt EB, Calhoun VD (2012). Data Visualization in the Neurosciences: overcoming the Curse of Dimensionality. Neuron 74:603-608.

Amrhein V, Greenland S, McShane B (2019). Scientists rise up against statistical significance. Nature 567:305–307

Anscombe FJ (1973). Graphs in Statistical Analysis. The American Statistician 27(1):17-21

Bacchetti P (2013). Small sample size is not the real problem. Nat Rev Neuroscience 14, 585.

Benjamin DJ, Berger JO, Johannesson M, Nosek BA, Wagenmakers EJ, Berk R, Bollen KA, Brembs B, Brown L, Camerer C, Cesarini D, Chambers CD, Clyde M, Cook TD, De Boeck P, Dienes Z, Dreber A, Easwaran K, Efferson C, Fehr E, Fidler F, Field AP, Forster M, George EI, Gonzalez R, Goodman S, Green E, Green DP, Greenwald AG, Hadfield JD, Hedges LV, Held L, Hua Ho T, Hoijtink H, Hruschka DJ, Imai K, Imbens G, Ioannidis JPA, Jeon M, Jones JH, Kirchler M, Laibson D, List J, Little R, Lupia A, Machery E, Maxwell SE, McCarthy M, Moore DA, Morgan SL, Munafó M, Nakagawa S, Nyhan B, Parker TH, Pericchi L, Perugini M, Rouder J, Rousseau J, Savalei V, Schönbrodt FD, Sellke T, Sinclair B, Tingley D, Van Zandt T, Vazire S, Watts DJ, Winship C, Wolpert RL, Xie Y, Young C, Zinman J, Johnson VE (2018). Redefine statistical significance. Nat Hum Behav 2(1):6-10.

Bennett CM, Baird AA, Miller MB, Wolford GL (2009). Neural correlates of interspecies perspective taking in the post-mortem Atlantic salmon: an argument for proper multiple comparisons correction. J Serendipitous Unexpected Results 1:1-5.

Bhayana R (2024). Chatbots and Large Language Models in Radiology: A Practical Primer for Clinical and Research Applications. Radiology 310(1):e232756.

Bishay S, Robb WH, Schwartz TM, Smith DS, Lee LH, Lynn CJ, Clark TL, Jefferson AL, Warner JL, Rosenthal EL, Murphy BA, Hohman TJ, Koran MEI (2024). Frontal and anterior temporal hypometabolism post chemoradiation in head and neck cancer: A real-world PET study. J Neuroimaging 34(2):211-216.

Biswal BB, Mennes M, Zuo XN, Gohel S, Kelly C, et al. (2010). Toward discovery science of human brain function. Proc Natl Acad Sci U S A 107(10):4734-9.

Botvinik-Nezer R, Holzmeister F, Camerer CF, Dreber A, Huber J, Johannesson M, et al. (2020). Variability in the analysis of a single neuroimaging dataset by many teams. Nature 582(7810):84-88.

Boutron I, Ravaud P (2018). Misrepresentation and distortion of research in biomedical literature. Proc Natl Acad Sci U S A 115(11):2613-2619.

Bowring A, Telschow F, Schwartzman A, Nichols TE (2019). Spatial confidence sets for raw effect size images. Neuroimage 203:116187. doi: 10.1016/j.neuroimage.2019.116187.

Bzdok D, Thieme A, Levkovskyy O, Wren P, Ray T, Reddy S (2024). Data science opportunities of large language models for neuroscience and biomedicine. Neuron 112(5):698-717. doi: 10.1016/j.neuron.2024.01.016.

Calhoun VD, Adali T, Pearlson GD, Pekar JJ (2001). A method for making group inferences from functional MRI data using independent component analysis. Hum Brain Mapp 14(3):140-51. doi: 10.1002/hbm.1048.

Calhoun VD, Pekar JJ, McGinty VB, Adali T, Watson TD, Pearlson GD (2002). Different activation dynamics in multiple neural systems during simulated driving. Hum Brain Mapp 16(3):158-67. doi: 10.1002/hbm.10032.

Chen G, Taylor PA, Cox RW (2017). Is the statistic value all we should care about in neuroimaging? Neuroimage. 147:952-959.

Chen G, Taylor PA, Qu X, Molfese PJ, Bandettini PA, Cox RW, Finn ES (2020). Untangling the relatedness among correlations, part III: Inter-subject correlation analysis through Bayesian multilevel modeling for naturalistic scanning. Neuroimage 216:116474. doi: 10.1016/j.neuroimage.2019.116474.

Chen G, Pine DS, Brotman MA, Smith AR, Cox RW, Haller SP (2021). Trial and error: A hierarchical modeling approach to test-retest reliability. Neuroimage 245:118647.

Chen G, Pine DS, Brotman MA, Smith AR, Cox RW, Taylor PA, Haller SP (2022). Hyperbolic trade-off: the importance of balancing trial and subject sample sizes in neuroimaging. NeuroImage 247:118786.

Chen G, Taylor PA, Stoddard J, Cox RW, Bandettini PA, Pessoa L (2022). Sources of information waste in neuroimaging: mishandling structures, thinking dichotomously, and over-reducing data. Aperture Neuro. 2: DOI: 10.52294/2e179dbf-5e37-4338-a639-9ceb92b055ea

Coursey SE, Mandeville J, Reed MB, Hartung GA, Garimella A, Sari H, Lanzenberger R, Price JC, Polimeni JR, Greve DN, Hahn A, Chen JE (2024). On the analysis of functional PET (fPET)-

FDG: baseline mischaracterization can introduce artifactual metabolic (de)activations. bioRxiv [Preprint] 2024.10.17.618550.

Cox RW (1996). AFNI: software for analysis and visualization of functional magnetic resonance neuroimages. Comput Biomed Res 29(3):162-173. doi:10.1006/cbmr.1996.0014

Cremers HR, Wager TD, Yarkoni T (2017). The relation between statistical power and inference in fMRI. PLoS One 12(11):e0184923.

Dafflon J, F Da Costa P, Váša F, Monti RP, Bzdok D, Hellyer PJ, Turkheimer F, Smallwood J, Jones E, Leech R (2022). A guided multiverse study of neuroimaging analyses. Nat Commun 13(1):3758. doi: 10.1038/s41467-022-31347-8.

Damoiseaux JS, Rombouts SA, Barkhof F, Scheltens P, Stam CJ, Smith SM, Beckmann CF (2006). Consistent resting-state networks across healthy subjects. Proc Natl Acad Sci U S A 103(37):13848-53.

Etzel JA, Zacks JM, Braver TS (2013). Searchlight analysis: promise, pitfalls, and potential. Neuroimage 78:261-9.

Fair DA, Cohen AL, Power JD, Dosenbach NU, Church JA, Miezin FM, Schlaggar BL, Petersen SE (2009). Functional brain networks develop from a "local to distributed" organization. PLoS Comput Biol 5(5):e1000381.

Fischl B, Dale AM (2000). Measuring the thickness of the human cerebral cortex from magnetic resonance images. Proc Natl Acad Sci U S A 97(20):11050-5.

Fisher RA (1925). Statistical Methods for Research Workers, Oliver and Boyd, Edinburgh.

Flournoy JC, Vijayakumar N, Cheng TW, Cosme D, Flannery JE, Pfeifer JH (2020). Improving practices and inferences in developmental cognitive neuroscience. Dev Cogn Neurosci 45:100807.

Forman SD, Cohen JD, Fitzgerald M, Eddy WF, Mintun MA, Noll DC (1995). Improved assessment of significant activation in functional magnetic resonance imaging (fMRI): use of a cluster-size threshold. Magn Reson Med 33(5):636-47.

Freund MC, Chen R, Chen G, Braver TS (2025). Complementary benefits of multivariate and hierarchical models for identifying individual differences in cognitive control. Imaging Neuroscience (2025) 3: imag_a_00447. https://doi.org/10.1162/imag_a_00447

Goebel R (2012). BrainVoyager--past, present, future. NeuroImage 62, 748–56. https://doi.org/10.1016/j.neuroimage.2012.01.083

Gonzalez-Castillo J, Saad ZS, Handwerker DA, Inati SJ, Brenowitz N, Bandettini PA (2012). Whole-brain, time-locked activation with simple tasks revealed using massive averaging and model-free analysis. Proc Natl Acad Sci U S A 109(14):5487-92.

Gorgolewski KJ, Storkey AJ, Bastin ME, Pernet CR (2012). Adaptive thresholding for reliable topological inference in single subject fMRI analysis. Front Hum Neurosci 6:245. doi: 10.3389/fnhum.2012.00245.

Gorgolewski KJ, Varoquaux G, Rivera G, Schwarz Y, Ghosh SS, Maumet C, Sochat VV, Nichols TE, Poldrack RA, Poline JB, Yarkoni T, Margulies DS (2015). NeuroVault.org: a web-based repository for collecting and sharing unthresholded statistical maps of the human brain. Front Neuroinform. 9:8. doi: 10.3389/fninf.2015.00008.

Greicius MD, Krasnow B, Reiss AL, Menon V (2003). Functional connectivity in the resting brain: a network analysis of the default mode hypothesis. Proc Natl Acad Sci U S A 100(1):253-8.

Hagmann P, Cammoun L, Gigandet X, Meuli R, Honey CJ, Wedeen VJ, Sporns O (2008). Mapping the structural core of human cerebral cortex. PLoS Biol 6(7):e159.

Halsey LG, Curran-Everett D, Vowler SL, Drummond GB (2015). The fickle P value generates irreproducible results. Nat Methods 12(3):179-85.

Hanayik T, Rorden C, Drake C, Thual A, Taylor P, Hardcastle N, McCarthy P, Androulakis A, Markiewicz C, Nedelec P (2023). niivue/niivue: 0.37.0 (0.37.0). Zenodo. 10.5281/zenodo.8320937

Ioannidis JP, Munafò MR, Fusar-Poli P, Nosek BA, David SP (2014). Publication and other reporting biases in cognitive sciences: detection, prevalence, and prevention. Trends Cogn Sci 18(5):235-41. doi: 10.1016/j.tics.2014.02.010.

Jennings RG, Van Horn JD (2012). Publication bias in neuroimaging research: implications for meta-analyses. Neuroinformatics 10(1):67-80. doi: 10.1007/s12021-011-9125-y.

Jernigan TL, Gamst AC, Fennema-Notestine C, Ostergaard AL (2003). More "mapping" in brain mapping: statistical comparison of effects. Hum Brain Mapp 19(2):90-5.

Jung B, Taylor PA, Seidlitz PA, Sponheim C, Perkins P, Ungerleider LG, Glen DR, Messinger A (2021). A Comprehensive Macaque FMRI Pipeline and Hierarchical Atlas. NeuroImage 235:117997.

Kennedy-Shaffer L (2019). Before *p* < 0.05 to Beyond *p* < 0.05: Using History to Contextualize *p*-Values and Significance Testing. Am Stat 73(Suppl 1):82-90.

Kinsey S, Kazimierczak K, Camazón PA, Chen J, Adali T, Kochunov P, Adhikari B, Ford J, van Erp TGM, Dhamala M, Calhoun VD, Iraji A (2024). Networks extracted from nonlinear fMRI connectivity exhibit unique spatial variation and enhanced sensitivity to differences between individuals with schizophrenia and controls. Nat Ment Health. 2024;2(12):1464-1475.

Lefort-Besnard J, Nichols TE, Maumet C (2024). Statistical Inference for Same Data Meta-Analysis in Neuroimaging Multiverse Analyzes. hal-04754078v2

Lerch J, Hammill C, van Eede M, Cassel D (2017). RMINC: Statistical Tools for Medical Imaging NetCDF (MINC) Files. R package version 1.5.2.1, http://mouse-imaging-centre.github.io/RMINC.

Lohmann G, Stelzer J, Müller K, Lacosse E, Buschmann T, Kumar VJ, Grodd W, Scheffler K (2017). Inflated false negative rates undermine reproducibility in task-based fMRI. bioRxiv 122788; doi: https://doi.org/10.1101/122788.

Luo WL, Nichols TE (2003). Diagnosis and exploration of massively univariate neuroimaging models. Neuroimage 19(3):1014-32.

Mandino F, Cerri DH, Garin CM, Straathof M, van Tilborg GAF, Chakravarty MM, Dhenain M, Dijkhuizen RM, Gozzi A, Hess A, Keilholz SD, Lerch JP, Shih YI, Grandjean J (2020). Animal Functional Magnetic Resonance Imaging: Trends and Path Toward Standardization. Front Neuroinform 13:78.

Maumet C, Auer T, Bowring A, Chen G, Das S, Flandin G, Ghosh S, Glatard T, Gorgolewski KJ, Helmer KG, Jenkinson M, Keator DB, Nichols BN, Poline JB, Reynolds R, Sochat V, Turner J, Nichols TE (2016). Sharing brain mapping statistical results with the neuroimaging data model. Sci Data 3:160102.

Maumet C, Nichols TE (2016). Minimal Data Needed for Valid and Accurate Image-Based fMRI Meta-Analysis. bioRxiv. doi: 10.1101/048249

McCarthy P (2024). FSLeyes (1.11.0). Zenodo. https://doi.org/10.5281/zenodo.11047709

Nichols T, Hayasaka S (2003). Controlling the familywise error rate in functional neuroimaging: a comparative review. Stat Methods Med Res 12(5):419-46.

Nichols TE, Das S, Eickhoff SB, Evans AC, Glatard T, Hanke M, Kriegeskorte N, Milham MP, Poldrack RA, Poline JB, Proal E, Thirion B, Van Essen DC, White T, Yeo BT (2017). Best practices in data analysis and sharing in neuroimaging using MRI. Nat Neurosci 20(3):299-303.

Nilearn contributors (2025) 'nilearn'. Zenodo. doi: 10.5281/zenodo.14697221.

Noble S, Curtiss J, Pessoa L, Scheinost D (2024). The tip of the iceberg: A call to embrace anti-localizationism in human neuroscience research. Imaging Neuroscience 2: 1–10.

NPG (2018). Checklists work to improve science. Nature 556:273–274. doi: 10.1038/d41586-018-04590-7.

Pang JC, Aquino KM, Oldehinkel M, Robinson PA, Fulcher BD, Breakspear M, Fornito A (2023). Geometric constraints on human brain function. Nature 618(7965):566-574.

Pernet CR, Madan CR (2020). Data visualization for inference in tomographic brain imaging. Eur J Neurosci 51(3):695-705.

Pessoa L (2014). Understanding brain networks and brain organization. Phys Life Rev 11(3):400-35.

Posse S, Wiese S, Gembris D, Mathiak K, Kessler C, Grosse-Ruyken M, Elghahwagi B, Richards T, Dager S, Kiselev V (1999). Enhancement of BOLD-contrast sensitivity by single-shot multi-echo functional MR imaging. Magnetic Resonance in Medicine. 42:87–97.

Reynolds RC, Taylor PA, Glen DR (2023). Quality control practices in FMRI analysis: Philosophy, methods and examples using AFNI. Front. Neurosci. 16:1073800. doi: 10.3389/fnins.2022.1073800

Reynolds RC, Glen DR, Chen G, Saad ZS, Cox RW, Taylor PA (2024). Processing, evaluating and understanding FMRI data with afni_proc.py. Imaging Neuroscience 2:1-52.

Ruff IM, Petrovich Brennan NM, Peck KK, Hou BL, Tabar V, Brennan CW, Holodny AI (20008). Assessment of the language laterality index in patients with brain tumor using functional MR imaging: effects of thresholding, task selection, and prior surgery. AJNR Am J Neuroradiol 29(3):528-35.

Russ BE, Leopold DA (2015). Functional MRI mapping of dynamic visual features during natural viewing in the macaque. Neuroimage 109:84-94.

Saad ZS, Reynolds RC, Argall B, Japee S, Cox RW (2004). SUMA: an interface for surface-based intra- and inter-subject analysis with AFNI. Presented at the 2nd IEEE International Symposium on Biomedical Imaging: Nano to Macro (IEEE Cat No. 04EX821), pp. 1510-1513 Vol. 2.

Saad ZS, Reynolds RC (2012). SUMA. Neuroimage 62, 768–773.

Sava-Segal C, Grall, C Finn ES (2024). Narrative 'twist' shifts within-individual neural representations of dissociable story features. bioRxiv preprint. https://www.biorxiv.org/content/10.1101/2025.01.13.632631v1

Savoy RL (2001). History and future directions of human brain mapping and functional neuroimaging. Acta Psychol (Amst) 107(1-3):9-42.

Seghier ML (2008). Laterality index in functional MRI: methodological issues. Magn Reson Imaging 26(5):594-601.

Shakespeare W (1597). An Excellent Conceited Tragedie of Romeo and Juliet. John Danter (London).

Shapiro, FR (2006). The New Yale Book of Quotations. Yale University Press (New Haven), p231.

Smith SM, Jenkinson M, Woolrich MW, Beckmann CF, Behrens TE, Johansen-Berg H, Bannister PR, De Luca M, Drobnjak I, Flitney DE, Niazy RK, Saunders J, Vickers J, Zhang Y, De Stefano N, Brady JM, Matthews PM (2004). Advances in functional and structural MR image analysis and implementation as FSL. Neuroimage. 23 Suppl 1:S208-19.

Smith SM, Fox PT, Miller KL, Glahn DC, Fox PM, Mackay CE, Filippini N, Watkins KE, Toro R, Laird AR, Beckmann CF (2009). Correspondence of the brain's functional architecture during activation and rest. Proc Natl Acad Sci U S A 106(31):13040-5.

Smith SM, Nichols TE (2009). Threshold-free cluster enhancement: addressing problems of smoothing, threshold dependence and localisation in cluster inference. Neuroimage 44(1):83-98.

Smith AR, White LK, Leibenluft E, McGlade AL, Heckelman AC, Haller SP, Buzzell GA, Fox NA, Pine DS (2020). The heterogeneity of anxious phenotypes: neural responses to errors in treatment-Seeking anxious and behaviorally inhibited youths. Journal of the American Academy of Child & Adolescent Psychiatry 59(6):759-769.

Suarez RO, Whalen S, Nelson AP, Tie Y, Meadows ME, Radmanesh A, Golby AJ (2009). Threshold-independent functional MRI determination of language dominance: a validation study against clinical gold standards. Epilepsy Behav 16(2):288-97.

Sundermann B, Pfleiderer B, McLeod A, Mathys C (2024). Seeing more than the Tip of the Iceberg: Approaches to Subthreshold Effects in Functional Magnetic Resonance Imaging of the Brain. Clin Neuroradiol 34(3):531-539.

Szucs D, Ioannidis JP (2020). Sample size evolution in neuroimaging research: An evaluation of highly-cited studies (1990-2012) and of latest practices (2017-2018) in high-impact journals. Neuroimage 221:117164.

Taylor PA, Gotts SJ, Gilmore AW, Teves J, Reynolds RC (2022). A multi-echo FMRI processing demo including TEDANA in afni_proc.py pipelines. Proc. OHBM-2022. https://afni.nimh.nih.gov/pub/dist/OHBM2022/OHBM2022_tayloretal_apmulti.pdf

Taylor PA, Reynolds RC, Calhoun V, Gonzalez-Castillo J, Handwerker DA, Bandettini PA, Mejia AF, Chen G (2023). Highlight Results, Don't Hide Them: Enhance interpretation, reduce biases and improve reproducibility. Neuroimage 274:120138.

Taylor PA, Glen DR, Chen G, Cox RW, Hanayik T, Rorden C, Nielson DM, Rajendra JK, Reynolds RC (2024). A Set of FMRI Quality Control Tools in AFNI: Systematic, in-depth and interactive QC with afni_proc.py and more. Imaging Neuroscience 2: 1–39.

Turner BO, Paul EJ, Miller MB, Barbey AK (2018). Small sample sizes reduce the replicability of task-based fMRI studies. Communications biology 1(1):62.

Vanduffel W, Fize D, Mandeville JB, Nelissen K, Van Hecke P, Rosen BR, Tootell RB, Orban GA (2001). Visual motion processing investigated using contrast agent-enhanced fMRI in awake behaving monkeys. Neuron 32(4):565-77. doi: 10.1016/s0896-6273(01)00502-5.

Van Essen DC, Smith SM, Barch DM, Behrens TE, Yacoub E, Ugurbil K; WU-Minn HCP Consortium (2013). The WU-Minn Human Connectome Project: an overview. Neuroimage 80:62-79.

Visconti di Oleggio Castello M, Halchenko YO, Guntupalli JS, Gors JD, Gobbini MI (2017). The neural representation of personally familiar and unfamiliar faces in the distributed system for face perception. Sci Rep 7(1):12237. doi: 10.1038/s41598-017-12559-1.

Voets N, Ashtari M, Beckmann C, Benjamin C, Benzinger T, Binder JR, ... Bookheimer S (2025). Consensus recommendations for clinical functional MRI applied to language mapping. *Aperture Neuro* 5. doi:10.52294/001c.128149

Wasserstein RL Lazar NA (2016). The ASA Statement on p-Values: Context, Process, and Purpose. The American Statistician 70(2):129–133. doi: 10.1080/00031305.2016.1154108.

Wicherts JM, Veldkamp CLS, Augusteijn HEM, Bakker M, van Aert RCM, & van Assen MALM (2016). Degrees of freedom in planning, running, analyzing, and reporting psychological studies: A checklist to avoid p-hacking. Frontiers in Psychology, 7, Article 1832. doi:10.3389/fpsyg.2016.01832

Yarkoni T, Poldrack RA, Nichols TE, Van Essen DC, Wager TD (2011). Large-scale automated synthesis of human functional neuroimaging data. Nat Methods. 8(8):665-70.

**SUPPLEMENTS**

*Background: showing more results and transparent thresholding*

Suggestions for displaying subthreshold results in neuroimaging go back at least two decades. Jernigan et al. (2003) suggested colored bands of statistics, and Luo and Nichols (2003) created explorable montages that included unthresholded data. Allen et al. (2012) placed outlines around suprathreshold regions and then displayed the remaining results with transparency increasing as the magnitudes shrunk ("transparent thresholding"). Bowring et al. (2019) suggested colored bands of effect confidence intervals. Pernet and Madan (2020) proposed several ideas, including displaying unthresholded data. Sundermann et al. (2024) used both transparency and different color ranges.

While not yet the standard format for image creation, transparent thresholding has been usefully applied in many neuroscience studies. With the goal of helping researchers see how transparent thresholding has been used in practice, we list a subset of peer reviewed publications that have used transparent thresholding in Table S1. Most were found using GoogleScholar to reference Allen et al. (2012) and/or Taylor et al. (2023). Studies cover a wide range of designs for FMRI, structural MRI and even non-MRI modalities, as well as studying both human and nonhuman subjects. We note that even for FMRI, several of these studies all displayed effect estimates (instead of simply statistics) as overlay maps, which was also recommended in the main text for more informative results reporting.

| Reference | Comment |
|---|---|
| Sadagopan et al. (2015) | Task-based FMRI of marmosets at 7T, overlaying effect estimates |
| Russ and Leopold (2015) | Video watching FMRI of macaques at 4.7T, individual subject maps |
| Pollmann et al. (2016) | Task-based of reward, voxelwise maps, overlaying effect estimates |
| Axelrod (2016) | Task-based FMRI, voxelwise contrast maps, overlaying effect estimates |
| Guevara et al. (2017) | Resting state FMRI of newborn rats, connectivity with HbR contrast |
| Cohen et al. (2018) | Structural MRI, individual-to-group VBM, overlaying effect estimates |
| Arend et al. (2018) | Structural MRI, group-based VBM, overlaying effect estimates |
| van Lieshout et al. (2018) | Task-based FMRI, voxelwise contrast maps, overlaying effect estimates |
| Richter et al. (2018) | Task-based FMRI, voxelwise contrast maps, overlaying effect estimates |
| Ellingson et al. (2018) | Task-based FMRI, voxelwise contrast maps, overlaying effect estimates |
| Keidel et al. (2018) | Naturalistic FMRI, voxelwise contrast maps, overlaying effect estimates |
| Dojat et al. (2018) | Structural MRI, group-based VBM, overlaying effect estimate contrast |
| Ziontz et al. (2019) | $^{18}$F-AV-1451 PET, SUVR-based modeling, overlaying GLM coefficients |
| Martinez-Saito et al. (2019) | Task-based FMRI, voxelwise contrast maps |
| Ai et al. (2019) | Structural MRI, multiple regression analyses, overlaying effect estimates |
| Chaze et al. (2019) | MR elastography, stiffness maps |
| Polsek et al. (2020) | Structural MRI of rodents at 7T |
| Handwerker et al. (2020) | Resting state FMRI and theta-burst TMS, correlation changes |
| Hofmans et al. (2020) | Task-based FMRI, voxelwise coefficients, overlaying effect estimates |
| Dellert et al. (2021) | Simultaneous EEG-FMRI, PPMs, overlaying effect estimates |
| Brolsma et al. (2021) | Task-based FMRI, voxelwise GLMs, overlaying effect estimates |
| van den Bosch et al. (2022) | Task-based pharmaco-FMRI, overlaying effect estimates |
| Böttinger et al. (2022) | Task-based FMRI, voxelwise contrast maps, overlaying effect estimates |
| van den Bosch et al. (2023) | Task-based pharmaco-FMRI, overlaying effect estimates |
| Boulakis et al. (2023) | Resting state FMRI with experience sampling |
| Orwig et al. (2023) | Resting state FMRI with GLMs and network analysis |
| Bishay et al. (2024) | FDG-PET, assoc. of chemoradiation and change in glucose metabolism |
| Strigo et al. (2024) | Task-based FMRI, voxelwise association maps |
| Reddy, Zvolanek et al. (2024) | Task-based ME-FMRI, voxelwise contrasts, |
| FIorito et al. (2024) | Meta-analysis of FMRI, overlaying effect estimates (mean Hedges' g) |
| Kinsey et al. (2024) | Task-based FMRI, voxelwise contrast maps, overlaying effect estimates |
| Coursey et al. (2024) | FDG-PET and FMRI, various measures, surface-based analyses |
| Reddy, Clements, et al. (2024) | Task-based ME-FMRI, voxelwise contrasts, overlaying effect estimates |
| Beynel, et al. (2024) | Task-based FMRI, voxelwise contrasts, overlaying effect estimates |
| Avery, et al. (2025) | Task-based FMRI, voxelwise contrasts |
| Mantas et al. (2025) | Structural MRI of mice at 9.4T, TBM assoc., overlaying effect estimates |
| Freund et al. (2025) | Task-based FMRI, test-retest reliability, ROI-based surface analysis |
| Aloi et al. (2025) | Resting state FMRI, group differences in functional correlation |

**Table S1.** *A non-exhaustive list of studies using transparent thresholding in figures. The majority of these studies also display effect estimates as overlays, rather than just statistics, which provides useful information (as discussed in the main text).*
*VBM = voxel-based morphometry. PET = positron emission tomography. SUVR = standardized uptake value ratios. GLM = general linear model. TMS = transcranial magnetic stimulation. PPM = posterior probability maps. FDG = fluorodeoxyglucose. ME = multi-echo. TBM = tensor-based morphometry*

*FMRI data and processing*

Related data for the examples are freely available via OSF (https://osf.io/n4a37). Processing and visualization scripts are freely available via GitHub, with links provided for each case below.

The task-based FMRI data for Ex. 1 have been previously acquired and described; see Smith et al. (2020) and Chen et al. (2021) for details. Briefly, 42 healthy youth and adults were scanned while performing a modified Eriksen Flanker task with two stimulus classes, congruent and incongruent. Data were acquired in two separate sessions with a total of 8 runs, containing 432 trials for each of the two conditions (of which only trials with correct responses were used in the analysis). Echo-planar images (EPIs) had: flip angle=60°, TE=25 ms, TR=2000 ms, 170 volumes per run, voxel=2.5x2.5x3.0 mm$^3$. Accompanying structural anatomical was acquired at 1mm isotropic resolution using T1-weighted a standard magnetization-prepared rapid acquisition gradient echo (MPRAGE) sequence. The data were processed using AFNI (Cox, 1996) version 20.3.00. Variations of the processed data (shown in Fig. 1B) were created as follows: 3dcalc was used to negate and zero the left hemisphere (Panels 2-3, respectively); a combination of 3dClustSim and 3dcalc were used to create Panel 4.

The task-based FMRI data for Ex. 2 and 3 were previously acquired as part of the NARPS project (Botvinik-Nezer et al., 2020); see that paper for acquisition details. Briefly, two groups of 54 subjects each performed variations of a mixed gambling paradigm task while being scanned, responding to potential gains and losses at various levels; there were four runs per subject. The EPI data were acquired with: flip angle=68°, TE=30ms, TR=1000ms, multiband=4, parallel factor (iPAT)=2, voxel size=2.0x2.0x2.4 mm$^3$, 453 volumes per run. Each subject's accompanying structural T1-weighted MPRAGE dataset had 1 mm isotropic resolution.

The group data used in Ex. 2 are publicly available through NeuroVault (Gorgolewski et al. 2015). Participating teams uploaded either *t*- or *Z*-statistic maps, which were unthresholded. The original NARPS data collection had 54 subjects in each of the two groups. Taylor et al. (2023) reported having 47 and 52 subjects per group after processing and QC, and we estimated that most teams' results would also have approximately 50 subjects remaining per group. Thus, the resulting degrees of freedom (DFs) would be enough to apply the same threshold value to either statistical dataset and consider it approximately equivalent (Fig. 2).

The FMRI processing for data used in Ex. 3 is described in Taylor et al. (2023), with accompanying GitHub repository of scripts (https://github.com/afni/apaper_highlight_narps). Standard processing included FreeSurfer (Fischl and Dale, 2000) version 7.1.1 and AFNI (Cox, 1996) version 22.2.12, particularly using afni_proc.py for full FMRI processing and quality control (Reynolds et al., 2024; Taylor et al., 2024).

In Figs. 2 and 3, similarity matrices are estimated to summarize properties across datasets in the cases where strict thresholding is applied or not. When opaque thresholding is applied, the results are binarized maps of suprathreshold regions, and therefore the Dice coefficient is appropriate for quantifying overlaps. In more formal analyses, separate Dice calculations would be made for sets of positive and negative regions (e.g., see Taylor et al., 2023), where reduced and similar values were observed, respectively; for simplicity here, all regions were put into a Dice comparison. For continuous value maps that exist when thresholding is not applied, Pearson correlation is appropriate to quantify the patterns in the data. Both Dice and Pearson similarities are straightforward, clear and widely used for the respective kinds of data. Conveniently, they also both have a maximum of 1 and minimum magnitude of 0 (with Pearson offering separate information of anticorrelation). We note that there may be other ways to quantify similarity. For example, for the continuous values one might choose to weight the correlations by the magnitude of the statistics themselves, thus providing greater influence to the high-statistic values in a fashion that more closely mimics the "highlighting" visualization of transparent thresholding; one could even use the alpha values that are calculated in the transparent thresholding visualization as weights. Either of these might be reasonable as high-statistic regions are often of the greatest interest. Separately, Gorgolewski et al. (2012) proposed an iterative approach of thresholding and Dice estimation to find a threshold that maximizes similarity.

In Ex. 4, previously acquired data by Bennett et al. (2009) of the "dead salmon" were no longer available (a query was made to the study authors, but they could not find the data). Therefore, the sole image from the original publication was used (in Fig. 4A and 4B-left), and an altered version showing what the "after correction" results would look like with strict thresholding was derived from that (Fig. 4B-right).

To present a more complete discussion of these results, a new "dead salmon" dataset was acquired with a task-based FMRI paradigm using a General Electric MR-750 3T MRI scanner. These data were collected in accordance with all local regulations and facility stipulations. Two runs of EPI data from a single *ex vivo* salmon (in a container with fluorinert to remove air pockets and gauze to minimize motion from scanner vibrations) were acquired with: flip angle=75°, TE=25 ms, TR=2000 ms, 170 volumes per run, voxel=2.0x2.0x2.5 mm$^3$. The accompanying structural T1-weighted MPRAGE dataset had 0.625x0.625x1.0 mm$^3$ resolution. The presented task stimulus was a standard flashing checkerboard with a 10-second duration, and there were 20 trials total across the two runs. Standard processing included AFNI version 24.1.05, particularly using afni_proc.py for full FMRI processing and quality control (Reynolds et al., 2024; Taylor et al., 2024). The full processing and visualization scripts are available (https://github.com/afni/apaper_gofigure_salmon), and demo downloads of both the raw and fully processed (including APQC HTML reports) are available at the project's OSF page (https://osf.io/n4a37).

*Brain images in the main text: displaying overlays and applying transparent thresholding*

The brain images displayed in Figs. 1-4 of the main text were created using AFNI (Cox, 1996), specifically using the toolbox's @chauffeur_afni command line program. This program facilitates systematic image and montage generation, by making many AFNI GUI and environment features scriptable, which is useful for figure generation and code sharing. The variously thresholded and faded colorbars were edited using AFNI's colorbar_tool.py program.

When displaying overlay data, AFNI allows for separating the dataset that is displayed (the "overlay" dataset, whose values are represented by colorbar mapping) from that which is thresholded (the "threshold" dataset, whose values are used to determine whether a voxel's overlay value will be displayed, or with what opacity in the case of transparent thresholding). For example, with opaque thresholding, a voxel's overlay dataset value would only be displayed if the corresponding threshold dataset value were suprathreshold; otherwise it would not be displayed. This partitioning of properties allows for the useful and common case of displaying effect estimates as overlay coloration while thresholding based on statistical information.

While it is still unfortunately common practice in the neuroimaging field to ignore effect estimate values and use statistics for both the overlay coloration and threshold, the benefits of visualizing the distinct effect information in results has been shown (Chen et al., 2017). The latter has been used here in Figs. 3-4, where per-voxel scaling has been applied to transform the unitless EPI data into meaningful BOLD percent signal change, which was then displayed as the overlay (while the accompanying statistic was used for thresholding). Only statistical information is available for each team's results in the public NARPS data, so Fig. 2 could only use this for both the overlay and threshold data. In Fig. 1, only the statistic is also used for both to not distract unaccustomed readers to the main point of the example. In the future, we hope that displaying effect estimates becomes more common practice, providing more information to readers and facilitating meta-analyses.

To perform transparent thresholding, AFNI implements a similar formulation as Allen et al. (2012). The transparency of an overlay voxel is calculated from the chosen "threshold" dataset, defining what is commonly termed the "alpha opacity", which is a continuous value between 0 and 1, inclusively. For a threshold dataset voxel of magnitude M and a selected threshold T, we have the following basic implementation with a quadratic decrease of alpha by default: if M>=T, then alpha = 1; else, alpha = (M/T)**2; users can also select a linear decrease in opacity, alpha=M/T, within the AFNI GUI, environment, or @chauffeur_afni program. The alpha value is then applied as a simple mixing fraction of the underlay and overlay RGB color vectors: RGB_final = (1-alpha)*RGB_ulay + alpha*RGB_olay. Thus, if alpha=1, the overlay is opaque (no underlay coloration is seen), while for intermediate values the overlay color fades into the background. As opposed to simply applying no thresholding to the overlay, this helps highlight regions of highest statistical significance (or of whatever dataset is applied as the "threshold" volume) while still allowing near-threshold results to be appreciated.

As a further highlighting measure, users can also place an outline around the suprathreshold regions, again following (Allen et al., 2012). It is default to use a black line, but AFNI allows a wide variety of other colors to be selected, as well, to maximize suitability across varied applications. For example, white outlines were used in Figs. 3 and 4.

*Software implementations and visualization*

Figs. 5 and 6 in the main text provides example illustrations of transparent thresholding across various software packages and implementations. We briefly describe the data and any relevant comments here.

The colorbars in these (and other) figures typically represent the use of transparent thresholding by fading to a background grayscale color (light gray or black). The fading can be either a linear or quadratic function of subthreshold value, as chosen by the researcher and/or software package implementation. When the overlay and thresholding datasets are the same, the fading is applied along the colorbar gradient itself. When the overlay and thresholding datasets differ (e.g., using an effect estimate for the former and a statistic for the latter), then the fading is orthogonal to the color gradient. Most images also include a black or white boundary around the suprathreshold regions; some software have yet to implement this fully, but are in process on doing so.

**SUMA.** This example (Fig. 5) uses a preliminary implementation of transparent thresholding in SUMA's surface mapping visualization (Saad et al., 2004; Saad and Reynolds, 2012). The image displays single subject results from a task-based FMRI study on an inflated surface representation, where the underlay grayscale shows the sulcal and gyral patterns. The data and processing scripts are part of the publicly available AFNI Bootcamp teaching material

(https://afni.nimh.nih.gov/pub/dist/doc/htmldoc/background_install/bootcamp_stuff.html; https://www.youtube.com/c/afnibootcamp). Specifically, the script *s03.ap.surface* was used to run afni_proc.py (Reynolds et al., 2024) to process the data and project it onto a FreeSurfer-estimated (Fischl and Dale, 2000) surface mesh that has been standardized using AFNI's *@SUMA_Make_Spec_FS*. The task FMRI data consisted of three runs of 150 time points each, during which a block design stimulus paradigm of "visual reliable" and "audio reliable" tasks were presented. The figure shows the full *F*-stat results (degrees of freedom: 2, 412) of the modeling, transparently thresholded at *F*=25. As with AFNI's implementation, the subthreshold fading of opacity is quadratic by default (as shown in Fig. 5), but can also be changed to linear.

**NiiVue.** This example (Fig. 5) uses the open source NiiVue (Hanayik et al. 2023) library to display single subject modeling results from a resting state study acquired with multi-echo (ME) FMRI data. NiiVue visualization is supported in all major web browsers, and enables application developers to create easily accessible web pages to publish data using these visualization techniques. This dataset is available as part of AFNI's ME-FMRI demo (downloadable via the command @Install_APMULTI_Demo1_rest; Taylor et al., 2022), and its acquisition and applications are described more in Gilmore et al. (2019) and in Gotts et al. (2020). The script *do_25_ap_me_br.tcsh* was used to run afni_proc.py (Reynolds et al., 2024) to process the data, including optimal combination of echos (Posse et al., 1999) as implemented within AFNI. By default, NiiVue uses an HTML canvas element to display volumetric data in a multiplanar layout with an optional 3D volume rendering tile. Each imaging plane can be navigated independently and volume rendering supports an interactive clip plane. The image displays Pearson correlation as for both overlay coloration and thresholding, with the transparent thresholding applied at |*r*|=0.3; NiiVue uses quadratic alpha fading, as in AFNI's default (described above). This panel is part of the afni_proc.py quality control (APQC) HTML (Taylor et al., 2024), which includes such toggleable Niivue instances for within-browser exploration (see an interactive online demo at https://afni.github.io/qc-demo-repo/). Additionally NiiVue's mesh visualization also contains transparent thresholding functionality for multilayered surface displays.

**AFNI.** This example (Fig. 5) uses AFNI (Cox, 1996) to display modeling results from a single macaque in task-based FMRI study. The data and scripts are available as part of a full processing demo (downloadable via the command @Install_MACAQUE_DEMO), which are described along with the acquisition paradigm in (Jung et al., 2021). The script *do_20_ap.tcsh* was used to run afni_proc.py (Reynolds et al., 2024) to process the data. The task data consisted of four runs of 112 time points each, during which a block design task of image presentation (faces, objects, scrambled faces and scrambled objects) were presented; MION (monocrystalline iron oxide nanoparticle; Vanduffel et al., 2001) was also applied as a contrast agent, and this was accounted for during processing. The displayed image shows the "(intact images) - (scrambled images)" contrast as the overlay in units of BOLD % signal change and the associated *t*-statistic as the thresholding volume. Transparent thresholding is applied at |*t*|=3.3, corresponding to *p*=0.001. This montage is part of afni_proc.py's APQC HTML (Taylor et al., 2024), to facilitate systematic checks of data. Transparent thresholding is particularly helpful to see potential subthreshold artifacts that may be part of the data (Reynolds et al., 2023).

An example of using AFNI to visualize transparent thresholding in an ROI-based analysis is also provided (Fig. 7). The image shows results of Bayesian multilevel modeling with AFNI's RBA program (Chen et al., 2019), for the task-based FMRI analysis of Hypothesis 2 of the NARPS dataset. The effect estimate value in units of BOLD percent signal change per dollar (as the paradigm included a gambling-related task stimulus) is shown within each ROI of the utilized Glasser atlas (Glasser et al., 2016). The values used for thresholding are the statistical evidence $E_s$ from the Bayesian modeling, essentially the posterior probability $P^+$, scaled to be in range [-1, 1]. Thresholding was applied at $|E_s| > 0.95$, with suprathreshold regions outlined in black and transparency fading at a quadratic rate. This image was created using a new program in AFNI called chauffeur_map_rois, which wraps around @chauffeur_afni.

**FSLeyes.** This example (Fig. 5) uses FSLeyes (McCarthy, 2024) to display single-subject modeling results from a task FMRI study, acquired as part of the publicly available FSL Course data set (https://open.win.ox.ac.uk/pages/fslcourse/website/). During the experiment, words were presented at different frequencies. Sentences were presented one word at a time, at frequencies ranging from 50 words per minute (wpm) to 1250 wpm, and the participant simply had to read the words as they were presented. The displayed image shows the results from a *F*-test comparing all pairs of frequencies, which was therefore sensitive to brain regions that responded differently to different word frequencies. Cluster-based thresholding was used to identify significant regions, with a cluster-forming threshold of *Z*=3.1 and cluster significance threshold of *p*=0.05. The example shows axial slices from the corresponding Z-statistic image, with the black outline highlighting significant clusters. Voxels with a *Z*-value at or above 3.1 are fully opaque, whereas the opacity of voxels with *Z*-value below 3.1 is linearly modulated by the *Z* value.

This visualisation can be achieved in FSLeyes for any statistic from a FSL FEAT analysis (Smith et al., 2004) by:
1. loading the un-thresholded and thresholded statistic images (e.g. *stats/zfstat1.nii.gz* and *thresh_zfstat1.nii.gz*), along with a suitable background image (e.g. *example_func.nii.gz*).
2. For the thresholded image, setting the *Overlay type* to *Mask*, and enabling the *Show outline only* option.
3. For the un-thresholded image, disabling the *Link display/clipping ranges*, and enabling the *Modulate alpha by intensity* option.
4. For the un-thresholded image, adjusting the positive and negative colour maps, and display and modulate range, as desired.

FSLeyes also allows the transparency of one image to be modulated by that of another image via the *Modulate by* option. This option allows one to, for instance, display contrast of parameter estimate (COPE) values and have their transparency modulated by the corresponding *Z*-statistics.

**Trends-Matlab & GIFT.** This example (Fig. 6) uses Matlab-based scripts (see https://trendscenter.org/x/datavis; Allen et al., 2012) to visualize task-based FMRI group results from a study by Kinsey et al. (2024). Comparisons between explicitly nonlinear (ENL) subject-level temporal (TEMP) and posterior default mode (pDM) intrinsic connectivity network (ICN) estimates derived from healthy controls (HC) and individuals with schizophrenia (SZ). Results are plotted according to a dual-coded colormap with transparency reflecting two-sided independent-samples *t*-statistic magnitudes (n = 508; df = 506), and contours indicate false discovery rate-corrected statistical significance ($q < 0.05$), as well as the outer edge of the masked EPI data. Warmer hues indicate HC > SZ, and cooler hues indicate SZ > HC. Results are overlaid on the ch2bet template with x, y and z coordinates listed relative to the origin in Montreal Neurological Institute 152 space. This thresholding approach is also implemented in the group ICA of fMRI toolbox (GIFT; http://trendscenter.org/software/gift).

**BrainVoyager.** The possibility to use transparent thresholding is available as a standard feature in BrainVoyager (Goebel, 2012). Contours of the thresholded map can be created using the 'Convert Map Clusters to VOI(s)' option. The upcoming version 24.0 release of BrainVoyager will perform the contouring as default when turning on transparent map thresholding. The displayed example (Fig. 6) uses the NeWBI4fMRI tutorial 1 dataset (see https://www.newbi4fmri.com/tutorial-1-data), which is a single participant's data. The task-based FMRI stimulation consisted of faces, hands, bodies, scrambled images, and fixation baseline conditions. The transparent thresholding example visual is created using 'General Linear Model: Single Study' with default parameters as implemented in BrainVoyager version 24, and it displays all stimulation conditions versus the fixation baseline contrast. The transparent coloring overlaid on the greyscale anatomical image reflects the statistical values below the chosen threshold q(FDR) < 0.05. The opaque black lines represent the statistical values above the q(FDR) < 0.05 threshold only at the borders of the spatial clusters.

**CIVET & minc-toolkit-v2.** The image (Fig. 6) was created with colour_object/ray_trace from the minc-toolkit-v2 (https://github.com/BIC-MNI/minc-toolkit-v2). Vertex-wise cortical thickness was estimated using CIVET (Ad-Dab'bagh et al., 2006), comparing the effect of first episode psychosis group vs a control group, with age and sex covariates. (The datasets come from unpublished work from the Prevention and Early Intervention for Psychosis (PEPP) clinic in Montreal, Canada.) Both the colormap and threshold volume are the *t*-statistic from the modeling. An arbitrary threshold of $|t| = 2$ has been applied, below which transparent thresholding occurs with a linear fade. No boundary contour has been placed around the suprathreshold regions, but there are plans to add this functionality.

**RMINC & MRIcrotome.** The image in Fig. 6 was created in the R-programming language, by using RMINC (Lerch et al., 2017) and the related MRIcrotome (https://github.com/Mouse-Imaging-Centre/MRIcrotome). The structural MRI data from Wistar rats were analyzed using deformation-based morphometry (DBM), as part of a study of the longitudinal effects of morphine self-administration (n = 33) versus a control group (n = 36), from postnatal day 60 (P60) to 81 (P81). The maps are the result of a 2x3 linear mixed model, for the interaction effect (Group [Morphine vs Control] x Time [T1,T2,T3]). The overlaid colormap describes the direction of the *t*-statistics: cool colors denote negative values compared to the control group, and warmer colors denote positive values. The transparent threshold is set at $|t| = 2.5$ (black line), which is the adjusted value for FDR = 5%, and fades linearly with decreasing statistic. Locations with even more significant statistic values are additionally highlighted with a yellow boundary (|*t*| = 3.1, which is where FDR = 1%).

An ROI-based example with RMINC visualization is also provided (Fig. 7). The ROI morphometry compares two inbred mouse strains (C57BL/6J vs DBA) and shows *t*-statistics encoded by both color and alpha-based transparency. The FDR-based threshold at 5% is indicated by solid black lines.

**Nilearn.** The transparency thresholding feature will be available in Nilearn (Nilearn contributors, 2025) from release 0.12 (scheduled for April 2025). It can be used by passing a statistical map in Nifti format (Cox et al., 2004) to the "*transparency*" parameter in several plotting functions in Nilearn such as: *plot_glass_brain*, *plot_stat_map* and *plot_img*. Users can use the "*transparency_range*" parameter to specify the range of values between which the transparency would vary. This would make the voxels with values below this range to be fully transparent and the ones above this range to be fully opaque. Fig. 7 shows a *t*-statistical contrast map from Thirion et al. (2014) (https://neurovault.org/images/10426/) overlaid on top of an MNI ICBM152 volumetric template via the *plot_stat_map* function. The full demonstration of this feature with other data and plotting functions is available on the development version of the Nilearn documentation (https://nilearn.github.io/dev/auto_examples/01_plotting/plot_transparency.html).

**bidspm.** The transparent plot in Fig. 7 is created using "Slice Display" (Zandbelt, 2017) implemented in bidspm (v4.0.0 - https://github.com/cpp-lln-lab/bidspm), a Matlab/Octave toolbox to perform MRI data analyses on a BIDS dataset (Gorgolewski et al., 2016) using SPM12 (Wellcome Trust Centre for Neuroimaging London, UK). The example shown displays the results obtained with the "stats" and "results" steps processing a single subject for the "listening" condition from the "Mother of All Experiments (MoAE)" dataset (https://www.fil.ion.ucl.ac.uk/spm/download/data/MoAEpilot). The transparent plot shows the *t*-statistic contrast map overlaid onto the normalized T1w structural image (IXI549Space), the transparent thresholding is applied at |*t*| = 5.3 corresponding to a voxelwise FWE-corrected threshold of *p* = 0.05 fading with decreasing statistic*,* and the black line represents the contour of the suprathreshold regions. Parameters such as image plane, number of slices, result maps, and applied transparency are fully customizable by the user. A demonstration of how to implement a transparent plot from bidspm results is available from the toolbox documentation (https://bidspm.readthedocs.io/en/latest/demos/moae.html). Future releases of bidspm will implement a fully automated generation of transparent plots from the results.

**ADDITIONAL REFERENCES IN SUPPLEMENTS**

Ai H, Xin Y, Luo YJ, Gu R, Xu P (2019). Volume of motor area predicts motor impulsivity. Eur J Neurosci 49(11):1470-1476. doi: 10.1111/ejn.14339. Epub 2019 Jan 29. PMID: 30636081.

Aloi J, Korin TE, Murray OK, Crum KI, LeFevre K, Dzemidzic M, Hulvershorn LA (2025). Latent Profiles of Impulsivity and Emotion Regulation in Children with Externalizing Disorders are Associated with Alterations in Striatocortical Connectivity. Biological Psychiatry: Cognitive Neuroscience and Neuroimaging (in press). https://doi.org/10.1016/j.bpsc.2025.02.013

Arend I, Yuen K, Sagi N, Henik A (2018). Neuroanatomical basis of number synaesthesias: A voxel-based morphometry study. Cortex 101:172-180. doi: 10.1016/j.cortex.2018.01.020. Epub 2018 Feb 7. PMID: 29482015.

Avery JA, Carrington M, Ingeholm JE, Darcey V, Simmons WK, Hall KD, Martin A. Automatic engagement of limbic and prefrontal networks in response to food images reflects distinct information about food hedonics and inhibitory control. Commun Biol. 2025 Feb 20;8(1):270. doi: 10.1038/s42003-025-07704-w. PMID: 39979602; PMCID: PMC11842766.

Axelrod V (2016). On the domain-specificity of the visual and non-visual face-selective regions. Eur J Neurosci 44(4):2049-63. doi:10.1111/ejn.13290. Epub 2016 Jul 1. PMID: 27255921.

Beynel L, Gura H, Rezaee Z, Ekpo EC, Deng ZD, Joseph JO, Taylor P, Luber B, Lisanby SH. Lessons learned from an fMRI-guided rTMS study on performance in a numerical Stroop task. PLoS One. 2024 May 6;19(5):e0302660. doi: 10.1371/journal.pone.0302660. PMID: 38709724; PMCID: PMC11073721.

Böttinger BW, Baumeister S, Millenet S, Barker GJ, Bokde ALW, Büchel C, Quinlan EB, Desrivières S, Flor H, Grigis A, Garavan H, Gowland P, Heinz A, Ittermann B, Martinot JL, Martinot MP, Artiges E, Orfanos DP, Paus T, Poustka L, Fröhner JH, Smolka MN, Walter H, Whelan R, Schumann G, Banaschewski T, Brandeis D, Nees F; IMAGEN Consortium (2022). Orbitofrontal control of conduct problems? Evidence from healthy adolescents processing negative facial affect. Eur Child Adolesc Psychiatry 31(8):1-10. doi: 10.1007/s00787-021-01770-1.

Boulakis PA, Mortaheb S, van Calster L, Majerus S, Demertzi A (2023). Whole-Brain Deactivations Precede Uninduced Mind-Blanking Reports. J Neurosci 43(40):6807-6815. doi: 10.1523/JNEUROSCI.0696-23.2023. Epub 2023 Aug 29. PMID: 37643862; PMCID: PMC10552942.

Brolsma SCA, Vassena E, Vrijsen JN, Sescousse G, Collard RM, van Eijndhoven PF, Schene AH, Cools R (2021). Negative Learning Bias in Depression Revisited: Enhanced Neural Response to Surprising Reward Across Psychiatric Disorders. Biol Psychiatry Cogn Neurosci

Neuroimaging 6(3):280-289. doi: 10.1016/j.bpsc.2020.08.011. Epub 2020 Aug 30. PMID: 33082119

Chaze CA, McIlvain G, Smith DR, Villermaux GM, Delgorio PL, Wright HG, Rogers KJ, Miller F, Crenshaw JR, Johnson CL (2019). Altered brain tissue viscoelasticity in pediatric cerebral palsy measured by magnetic resonance elastography. Neuroimage Clin. 2019;22:101750. doi: 10.1016/j.nicl.2019.101750. Epub 2019 Mar 7. PMID: 30870734; PMCID: PMC6416970.

Chen G, Xiao Y, Taylor PA, Rajendra JK, Riggins T, Geng F, Redcay E, Cox RW (2019). Handling Multiplicity in Neuroimaging Through Bayesian Lenses with Multilevel Modeling. Neuroinformatics. 17(4):515-545. doi:10.1007/s12021-018-9409-6

Cohen ZZ, Arend I, Yuen K, Naparstek S, Gliksman Y, Veksler R, Henik A (2018). Tactile enumeration: A case study of acalculia. Brain Cogn 127:60-71. doi:10.1016/j.bandc.2018.10.001. Epub 2018 Oct 16. PMID: 30340181.

Coursey SE, Mandeville J, Reed MB, Hartung GA, Garimella A, Sari H, Lanzenberger R, Price JC, Polimeni JR, Greve DN, Hahn A, Chen JE (2024). On the analysis of functional PET (fPET)-FDG: baseline mischaracterization can introduce artifactual metabolic (de)activations. bioRxiv [Preprint]. 2024 Oct 21:2024.10.17.618550. doi: 10.1101/2024.10.17.618550. PMID: 39484579; PMCID: PMC11526866.

Cox RW, Ashburner J, Breman H, Fissell K, Haselgrove C, Holmes CJ, Lancaster JL, Rex DE, Smith SM, Woodward JB, Strother SC (2004). A (sort of) new image data format standard: NiFTI-1. Presented at the 10th Annual Meeting of the Organization for Human Brain Mapping.

Dellert T, Müller-Bardorff M, Schlossmacher I, Pitts M, Hofmann D, Bruchmann M, Straube T (2021). Dissociating the Neural Correlates of Consciousness and Task Relevance in Face Perception Using Simultaneous EEG-fMRI. J Neurosci 41(37):7864-7875. doi: 10.1523/JNEUROSCI.2799-20.2021.

Dojat M, Pizzagalli F, Hupé JM (2018). Magnetic resonance imaging does not reveal structural alterations in the brain of grapheme-color synesthetes. PLoS One 13(4):e0194422.

Ellingson LD, Stegner AJ, Schwabacher IJ, Lindheimer JB, Cook DB (2018). Catastrophizing Interferes with Cognitive Modulation of Pain in Women with Fibromyalgia. Pain Med 19(12):2408-2422. doi: 10.1093/pm/pny008. PMID: 29474665; PMCID: PMC6659027.

Fiorito AM, Blasi G, Brunelin J, Chowdury A, Diwadkar VA, Goghari VM, Gur RC, Kwon JS, Quarto T, Rolland B, Spilka MJ, Wolf DH, Yun JY, Fakra E, Sescousse G (2024). Blunted brain responses to neutral faces in healthy first-degree relatives of patients with schizophrenia: an image-based fMRI meta-analysis. Schizophrenia 10(1):38.

Freund MC, Chen R, Chen G, Braver TS. Complementary benefits of multivariate and hierarchical models for identifying individual differences in cognitive control. Imaging Neurosci (Camb). 2025 Feb 10;3:imag_a_00447. doi: 10.1162/imag_a_00447. PMID: 39957839; PMCID: PMC11823007.

Gilmore AW, Kalinowski SE, Milleville SC, Gotts SJ, Martin A (2019). Identifying task-general effects of stimulus familiarity in the parietal memory network. Neuropsychologia 124:31-43. doi: 10.1016/j.neuropsychologia.2018.12.023.

Glasser MF, Coalson TS, Robinson EC, Hacker CD, Harwell J, Yacoub E, Ugurbil K, Andersson J, Beckmann CF, Jenkinson M, Smith SM, Van Essen DC (2016). A multi-modal parcellation of human cerebral cortex. Nature 536(7615):171-178.

Gorgolewski KJ, Auer T, Calhoun VD, Craddock RC, Das S, Duff EP, Flandin G, Ghosh SS, Glatard T, Halchenko YO, Handwerker DA, Hanke M, Keator D, Li X, Michael Z, Maumet C, Nichols BN, Nichols TE, Pellman J, Poline JB, Rokem A, Schaefer G, Sochat V, Triplett W, Turner JA, Varoquaux G, Poldrack RA (2016). The brain imaging data structure, a format for organizing and describing outputs of neuroimaging experiments. Sci Data 3:160044. doi: 10.1038/sdata.2016.44.

Gotts SJ, Gilmore AW, Martin A (2020). Brain networks, dimensionality, and global signal averaging in resting-state fMRI: Hierarchical network structure results in low-dimensional spatiotemporal dynamics. Neuroimage 2020 205:116289. doi: 10.1016/j.neuroimage.2019.116289.

Guevara E, Pierre WC, Tessier C, Akakpo L, Londono I, Lesage F, Lodygensky GA (2017). Altered Functional Connectivity Following an Inflammatory White Matter Injury in the Newborn Rat: A High Spatial and Temporal Resolution Intrinsic Optical Imaging Study. Front Neurosci 11:358. doi: 10.3389/fnins.2017.00358. PMID: 28725174; PMCID: PMC5495836.

Handwerker DA, Ianni G, Gutierrez B, Roopchansingh V, Gonzalez-Castillo J, Chen G, Bandettini PA, Ungerleider LG, Pitcher D (2020). Theta-burst TMS to the posterior superior temporal sulcus decreases resting-state fMRI connectivity across the face processing network. Netw Neurosci 4(3):746-760.

Hofmans L, van den Bosch R, Määttä JI, Verkes RJ, Aarts E, Cools R. The cognitive effects of a promised bonus do not depend on dopamine synthesis capacity. Sci Rep. 2020 Oct 5;10(1):16473. doi: 10.1038/s41598-020-72329-4. PMID: 33020514; PMCID: PMC7536197.

Keidel JL, Oedekoven CSH, Tut AC, Bird CM (2018). Multiscale Integration of Contextual Information During a Naturalistic Task. Cereb Cortex 28(10):3531-3539. doi: 10.1093/cercor/bhx218. PMID: 28968727.

Kinsey S, Kazimierczak K, Camazón PA, Chen J, Adali T, Kochunov P, Adhikari BM, Ford J, van Erp TGM, Dhamala M, Calhoun VD, Iraji A (2024). Networks extracted from nonlinear fMRI connectivity exhibit unique spatial variation and enhanced sensitivity to differences between individuals with schizophrenia and controls. Nat Ment Health 2(12):1464-1475. doi: 10.1038/s44220-024-00341-y. Epub 2024 Nov 21. PMID: 39650801; PMCID: PMC11621020.

Martinez-Saito M, Konovalov R, Piradov MA, Shestakova A, Gutkin B, Klucharev V (2019). Action in auctions: neural and computational mechanisms of bidding behaviour. Eur J Neurosci 50(8):3327-3348. doi: 10.1111/ejn.14492. Epub 2019 Jul 29. PMID: 31219633; PMCID: PMC6899836.

Mantas I, Flais I, Branzell N, Ionescu TM, Kim E, Zhang X, Cash D, Hengerer B, Svenningsson P. A molecular mechanism mediating clozapine-enhanced sensorimotor gating. Neuropsychopharmacology. 2025 Feb 11. doi: 10.1038/s41386-025-02060-z. Epub ahead of print. PMID: 39934408.

Orwig W, Diez I, Bueichekú E, Kelly CA, Sepulcre J, Schacter DL (2023). Intentionality of Self-Generated Thought: Contributions of Mind Wandering to Creativity. Creat Res J 35(3):471-480. doi: 10.1080/10400419.2022.2120286.

Polsek D, Cash D, Veronese M, Ilic K, Wood TC, Milosevic M, Kalanj-Bognar S, Morrell MJ, Williams SCR, Gajovic S, Leschziner GD, Mitrecic D, Rosenzweig I (2020). The innate immune toll-like-receptor-2 modulates the depressogenic and anorexiolytic neuroinflammatory response in obstructive sleep apnoea. Sci Rep 10(1):11475. doi: 10.1038/s41598-020-68299-2. PMID: 32651433; PMCID: PMC7351955.

Pollmann S, Eštočinová J, Sommer S, Chelazzi L, Zinke W (2016). Neural structures involved in visual search guidance by reward-enhanced contextual cueing of the target location. Neuroimage 124(Pt A):887-897. doi: 10.1016/j.neuroimage.2015.09.040. Epub 2015 Sep 30. PMID: 26427645.

Reddy NA, Zvolanek KM, Moia S, Caballero-Gaudes C, Bright MG (2024). Denoising task-correlated head motion from motor-task fMRI data with multi-echo ICA. Imaging Neurosci 2:10.1162/imag_a_00057. doi: 10.1162/imag_a_00057.

Reddy NA, Clements RG, Brooks JCW, Bright MG (2024). Simultaneous cortical, subcortical, and brainstem mapping of sensory activation. Cereb Cortex 34(6):bhae273. doi: 10.1093/cercor/bhae273.

Richter D, Ekman M, de Lange FP (2018). Suppressed Sensory Response to Predictable Object Stimuli throughout the Ventral Visual Stream. J Neurosci 38(34):7452-7461. doi:10.1523/JNEUROSCI.3421-17.2018. Epub 2018 Jul 20. PMID: 30030402; PMCID: PMC6596138.

Russ BE, Leopold DA (2015). Functional MRI mapping of dynamic visual features during natural viewing in the macaque. Neuroimage 109:84-94.

Sadagopan S, Temiz-Karayol NZ, Voss HU (2015). High-field functional magnetic resonance imaging of vocalization processing in marmosets. Sci Rep. 2015 Jun 19;5:10950. doi: 10.1038/srep10950. PMID: 26091254; PMCID: PMC4473644.

Strigo IA, Guerra SG, Torrisi S, Murphy E, Toor T, Goldman V, Alter BJ, Vu AT, Hecht R, Lotz J, Simmons AN, Mehling WE (2024). Enhancing chronic low back pain management: an initial neuroimaging study of a mobile interoceptive attention training. Front Pain Res 5:1408027. doi: 10.3389/fpain.2024.1408027. PMID: 39403233; PMCID: PMC11471628.

Thirion B, Varoquaux G, Grisel O, Poupon C, Pinel P (2014). Principal component regression predicts functional responses across individuals. Med Image Comput Comput Assist Interv. 17(Pt 2):741-8. doi: 10.1007/978-3-319-10470-6_92. PMID: 25485446.

van den Bosch R, Lambregts B, Määttä J, Hofmans L, Papadopetraki D, Westbrook A, Verkes RJ, Booij J, Cools R (2022). Striatal dopamine dissociates methylphenidate effects on value-based versus surprise-based reversal learning. Nat Commun 13(1):4962. doi: 10.1038/s41467-022-32679-1. PMID: 36002446; PMCID: PMC9402573.

van den Bosch R, Hezemans FH, Määttä JI, Hofmans L, Papadopetraki D, Verkes RJ, Marquand AF, Booij J, Cools R (2023). Evidence for absence of links between striatal dopamine synthesis capacity and working memory capacity, spontaneous eye-blink rate, and trait impulsivity. Elife 12:e83161. doi: 10.7554/eLife.83161. PMID: 37083626; PMCID: PMC10162803.

van Lieshout LLF, Vandenbroucke ARE, Müller NCJ, Cools R, de Lange FP (2018). Induction and Relief of Curiosity Elicit Parietal and Frontal Activity. J Neurosci 38(10):2579-2588. doi:10.1523/JNEUROSCI.2816-17.2018. Epub 2018 Feb 8. PMID: 29439166; PMCID: PMC6705901.

Zandbelt B (2017). Slice Display. figshare. 10.6084/m9.figshare.4742866

Ziontz J, Bilgel M, Shafer AT, Moghekar A, Elkins W, Helphrey J, Gomez G, June D, McDonald MA, Dannals RF, Azad BB, Ferrucci L, Wong DF, Resnick SM (2019). Tau pathology in cognitively normal older adults. Alzheimers Dement (Amst) 11:637-645. doi: 10.1016/j.dadm.2019.07.007. PMID: 31517026; PMCID: PMC6732758.